\documentclass[twocolumn,english]{IEEEtran}
\usepackage[T1]{fontenc}
\usepackage[latin9]{inputenc}
\usepackage{array}
\usepackage{booktabs}
\usepackage{amsthm}
\usepackage{amsmath}
\usepackage{amssymb}
\usepackage{graphicx}
\usepackage{esint}

\makeatletter

\providecommand{\tabularnewline}{\\}

\theoremstyle{plain}
\newtheorem{thm}{\protect\theoremname}
\theoremstyle{definition}
\newtheorem{defn}[thm]{\protect\definitionname}

\usepackage{babel}
\ifCLASSOPTIONcompsoc
\else
\fi
\usepackage{cite}\usepackage{MnSymbol}\usepackage{subfigure}

\usepackage{balance}

\newcommand{\E}{\mathop{\mathbb E}}
\let\labelindent\relax
\usepackage{enumitem}

\usepackage{babel}
\providecommand{\definitionname}{Definition}
\providecommand{\theoremname}{Theorem}

\makeatother

\usepackage{babel}
\providecommand{\definitionname}{Definition}
\providecommand{\theoremname}{Theorem}

\begin{document}

\title{Energy Consumption Of Visual Sensor Networks: Impact Of Spatio--Temporal
Coverage}

\author{Alessandro Redondi, Dujdow Buranapanichkit, Matteo Cesana, Marco
Tagliasacchi and~\\ Yiannis Andreopoulos%
\thanks{AR, MC and MT are with the Dipartimento
di Elettronica, Informazione e Bioingegneria, Politecnico di Milano,
P.zza Leonardo da Vinci, 32 - 20133, Milano - Italy, e-mail: firstname.lastname@polimi.it.
DB is with the Department of Electrical Engineering, Faculty of Engineering,
Prince of Songkla University, Hat Yai, Songkla, 90112 Thailand; email:
dujdow.b@psu.ac.th. YA is with the Electronic and Electrical Engineering
Department, University College London, Roberts Building, Torrington
Place, London, WC1E 7JE, UK; tel. +442076797303; fax. +442073889325;
email: i.andreopoulos@ucl.ac.uk. 

AR, MC and MT acknowledge the financial support of the Future and
Emerging Technologies (FET) programme within the Seventh Framework
Programme for Research of the European Commission, under FET-Open
grant number: 296676. YA was funded by the UK EPSRC, grant EP/K033166/1. 

This paper appears in the IEEE Trans. Circ. and Syst. for Video Technol., 2014. Copyright (c) 2014 IEEE. Personal use of this material is permitted.
However, permission to use this material for any other purposes must
be obtained from the IEEE by sending a request to pubs-permissions@ieee.org.  %
}}
\maketitle
\begin{abstract}
Wireless visual sensor networks (VSNs) are expected to play a major
role in future IEEE 802.15.4 personal area networks (PAN) under recently-established
collision-free medium access control (MAC) protocols, such as the
IEEE 802.15.4e-2012 MAC. In such environments, the VSN energy consumption
is affected by the number of camera sensors deployed (spatial coverage),
as well as the number of captured video frames out of which each node
processes and transmits data (temporal coverage). In this paper, we
explore this aspect for \emph{uniformly-formed} VSNs, i.e., networks
comprising identical wireless visual sensor nodes connected to a collection
node via a balanced cluster-tree topology, with each node producing
independent identically-distributed bitstream sizes after processing
the video frames captured within each network activation interval.
We derive analytic results for the energy-optimal spatio--temporal
coverage parameters of such VSNs under \emph{a-priori} known bounds
for the number of frames to process per sensor and the number of nodes
to deploy within each tier of the VSN. Our results are parametric
to the probability density function characterizing the bitstream size
produced by each node and the energy consumption rates of the system
of interest. Experimental results derived from a deployment of TelosB
motes under a collision-free transmission protocol and Monte-Carlo--generated
data sets reveal that our analytic results are always within 7\%
of the energy consumption measurements for a wide range of settings.
In addition, results obtained via a multimedia subsystem (BeagleBone
Linux Computer) performing differential Motion JPEG encoding and local
visual feature extraction from video frames show that the optimal
spatio--temporal settings derived by the proposed framework allow
for substantial reduction of energy consumption in comparison to \emph{ad-hoc}
settings. As such, our analytic modeling is useful for early-stage
studies of possible VSN deployments under collision-free MAC protocols
prior to costly and time-consuming experiments in the field.\end{abstract}

\section{Introduction}

The integration of low-power wireless networking technologies such
as IEEE 802.15.4-enabled transceivers with inexpensive camera hardware
has enabled the development of the so-called \emph{visual sensor networks}
(VSNs)\cite{charfi2009VSN}. VSNs can be thought of as networks of
wireless devices capable of sensing multimedia content, such as still
images and video, audio, depth maps, etc. Via the recent provisioning
of an all-IPv6 network layer under 6LoWPAN \cite{wang2012IPV6} and
the emergence of collision-free low-power medium access control (MAC)
protocols, such as the time slotted channel hopping (TSCH) of IEEE
802.15.4e-2012 \cite{tinka2010TSCH}, VSNs are expected to play a
major role in the Internet-of-Things (IoT) paradigm \cite{zorzi2010IoT,gubbi2013IoT}.



\begin{figure*}[t]
\centering{}

\includegraphics[bb=230bp 5bp 1500bp 540bp,clip,scale=0.33]{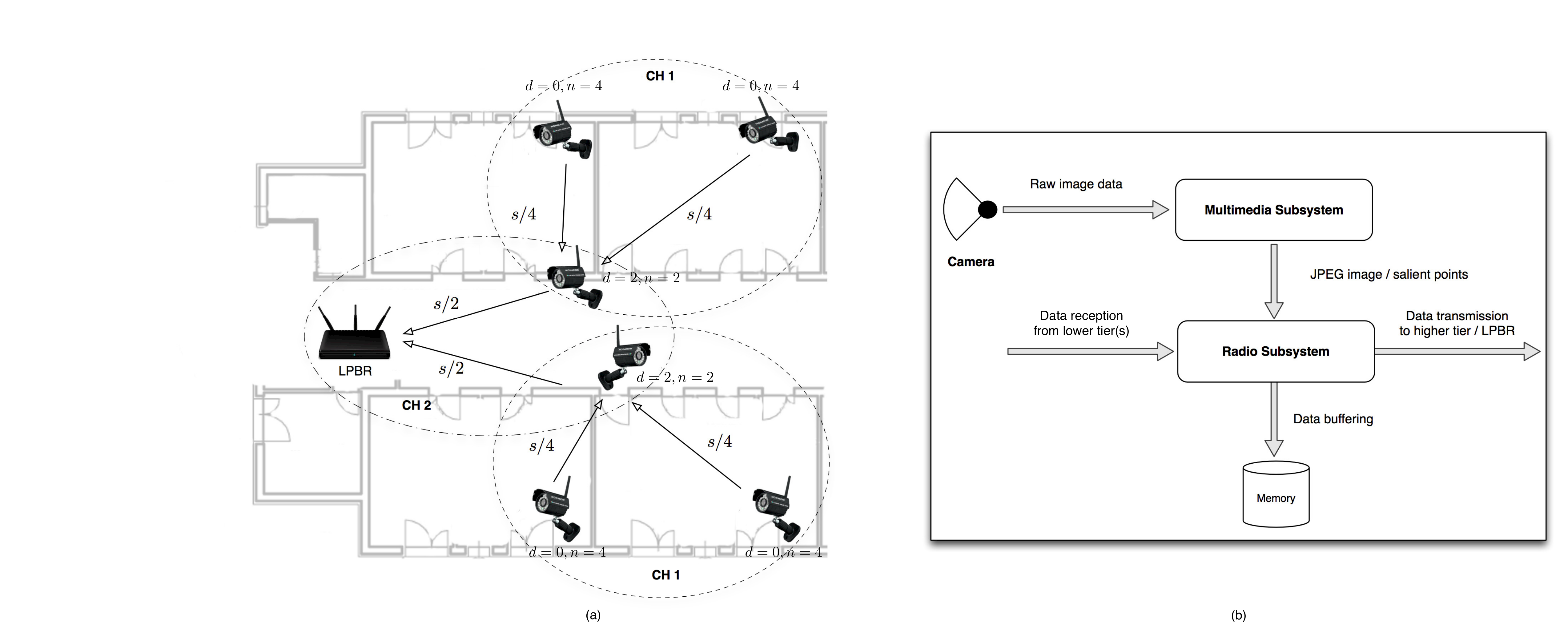}
\caption{(a) Two-tier uniformly-formed cluster-tree topology in a visual sensor
network for surveillance, where every visual sensor (video camera)
has its own spatial coverage (and different channels are used within
the indicated ellipses), with $s$ indicating the bits consumed by
each receiver/relay node within each active interval of $T$ seconds.
(b) Detail of the camera node system: each node comprises a multimedia
subsystem and a radio subsystem. If required, each node can buffer
parts of its data stream for later transmission. }

\label{fig:System_model} 
\end{figure*}

\subsection{Review of Visual Sensor Networks}

In comparison to traditional wireless sensor networks, VSNs are uniquely
challenging because of their heavy computational and bandwidth requirements
that stretch hardware and networking infrastructures to their limits.
Hence, an increasing number of VSN solutions were proposed recently,
focusing on: new transmission protocols allowing for high-bandwidth
collision-free communications \cite{ehsan2012survey}\cite{burana2012DTFDMA},
in-network processing techniques \cite{zuo2012two} and optimized
multimedia processing \cite{pudlewski2012compressed}. Also, several
hardware solutions have been proposed, with the aim of finding a VSN
platform that could be used for a broad range of multimedia tasks
\cite{mingorance2010efficient,newell2009self,CancliniSENSYS2013}.

Most of these proposed hardware solutions can be abstracted as two
tightly-coupled subsystems, shown in Figure \ref{fig:System_model}(b):
a multimedia processor board and a low-power radio subsystem \cite{citric,rahimi2005cyclops,hengstler2007mesheye},
interconnected via a push model. Within each node of the VSN, the
multimedia subsystem is responsible for acquiring images, processing
them and pushing the processed visual data to the radio subsystem,
which transmits it to a remote location. For example, in a traditional
surveillance application, the multimedia subsystem would compress
or process (e.g., extract visual features \cite{DBLP:journals/ijcv/Lowe04,DBLP:conf/MMSP/RedondiBCT})
the acquired images and push the resulting bitstream to the radio
subsystem for transmission to a central controller, where the data
would be analyzed or stored. 

Similar to traditional wireless sensor networks, VSN nodes are usually
battery operated. Hence, energy consumption plays a crucial role in
the design of a VSN, especially for those applications where a VSN
is required to operate for days or even weeks without external power
supply. In the last few years, several proposals strive for lifetime
maximization in VSNs. Specifically, solutions are available for energy-aware
protocols \cite{mingorance2010efficient,LeungMACEssentials}, cross-layer
optimization\cite{Tahir2013701}, application tradeoffs \cite{kandris2011energy}
and deployment strategies \cite{Munishwar:2013:CAV:2489253.2489262}.
While existing work addresses transmission, scheduling and protocol
design aiming for energy efficiency, it does not consider the impact
of the spatio--temporal coverage in the energy consumption of VSNs.
This is precisely the focus of this paper.

\subsection{Scenario}

\noindent We consider wireless visual sensor networks comprising a
cluster-tree topology, such as the one illustrated in Figure \ref{fig:System_model}(a),
where each camera node processes and transmits visual data to the
nodes of the higher tier, or to the Low-Power Border Router (LPBR)
\cite{wang2012IPV6} that can relay the streams to any IP address
over the Internet for analysis and processing. Moreover, we focus
on the case of a \emph{uniformly-formed} VSN, i.e. a network of identical
sensor nodes that, within each activation interval, are: \emph{(i)}
producing bitstream sizes with the same statistical characterization
and \emph{(ii)} connected to the base station via a balanced \emph{cluster-tree
topology }\cite{AlvesClusterTreeGTS}, represented by a symmetric
and acyclic graph with balanced bandwidth allocation per link. Each
node also relays streams stemming from $d$ other nodes of lower tier(s).
Within each node, the multimedia and radio subsystems work in parallel
{[}Figure \ref{fig:System_model}(b){]}: while the multimedia system
acquires and processes data corresponding to the current video frame,
the radio subsystem transmits (or relays) the multimedia stream stemming
from the processing of previous video frame(s).

\noindent Let $\frac{s}{T}$ kilobit-per-second (kbps) be the average
bandwidth at each node (in transmit or receive mode), with $s$ indicating
the bits consumed by each receiver/relay node over the VSN active
interval of $T$ seconds. For example, for a 802.15.4-compliant VSN
and $T=1$ second, the average consumption rate would be 250 kbps
at the physical layer. The MAC layer of the network is operating under
a collision-free time-division (or time-frequency division) multiple
access \cite{ehsan2012survey,burana2012DTFDMA,LeungMACEssentials,tinka2010TSCH,degesys2007desync},
so that each tier in the network can be configured in a way that simultaneous
transmissions in the same channel are avoided. The number of frames
captured by each camera during the operational time interval of the
VSN, i.e. each node's temporal coverage, is controlling the frequency
of the push operations. At the same time, the multimedia processing
task itself (e.g., image/video compression or extraction of visual
features) controls the size of the bitstream pushed to the radio subsystem
within each frame's duration. On the other hand, the number of sensors
in the same tier of the cluster-tree topology, i.e., the VSN's spatial
coverage, and the number of nodes whose bitstreams must be relayed
by each node (if any) control the bandwidth available to each sensor
(i.e., its average transmission rate) in each tier under a collision-free
MAC protocol. Therefore, there is a fundamental tradeoff between the
spatial and temporal coverage in a network: a large number of frames
leads to high bandwidth requirement per transmitter, which in turn
decreases the number of sensors that can be accommodated within each
tier of the VSN. Conversely, dense spatial coverage via the use of
a large number of visual sensors per tier decreases the available
bandwidth per sensor, which reduces the number of frames per sensor.

\subsection{Contribution and Paper Organization}

In this paper, we derive analytic results concerning energy-aware
VSN design under the push model of Figure \ref{fig:System_model}.
Specifically, we are interested in the link of the aforementioned
spatio--temporal tradeoff with the incurred energy consumption under
well-known probability density functions modeling the pushed bitstream
size of image and video applications, such as intra/inter-frame video
coding and local visual features extraction and transmission, and
make the following contributions:

\begin{itemize}[leftmargin=*]
\item

We derive an analytic model that captures the expected energy consumption
in function of: \emph{(i)} the number of visual sensors deployed at
each tier of the cluster-tree topology, \emph{(ii)} the number of
frames captured by each camera sensor within the operational time
interval and \emph{(iii)} the statistical characterization of the
bitstream data volume produced by each sensor after on-board multimedia
processing.

\item

The extrema of the derived energy consumption function are then analytically
derived in order to provide closed-form expressions for the minimum
energy consumption of each case under consideration.

\item

The analytic results are validated within two applications: video
coding and transmission based on differential Motion-JPEG and visual
feature extraction and transmission.

\end{itemize}

While our results are directly applicable to uniformly-formed VSNs,
we also indicate how they can be extended to non-uniformly formed
VSNs with varying statistical characterizations for the bitstream
sizes of different sensors and unbalanced bandwidth allocation for
the various links of each VSN tier during each activation interval. 

The rest of this paper is organized as follows: Section \ref{sec:System Model}
presents the proposed system model, while Section \ref{sec:min_energy_analysis}
presents the theoretical results; Section \ref{sec:Evaluation-of-Energy}
presents real-world experiments that validate the proposed framework
under controlled data production from each sensor, while Section \ref{sec:Applications}
presents results showcasing the accuracy of the proposed model under
real VSN data; finally, Section \ref{sec:Conclusions} concludes the
paper.

\section{Proposed System Model and Its Expected Energy Consumption\label{sec:System Model}}

In the following sections we introduce the components of the proposed
system model. The corresponding nomenclature is summarized in Table
\ref{tab:Nomenclature-table.1}. This sets the context for the derivation
of the expected energy consumption of each node of the uniformly-formed
visual sensor network in function of the utilized spatio--temporal
coverage settings.

\subsection{Spatio--Temporal Coverage and Statistical Characterization of Bitstream
Size per VSN Node}

We consider that the visual sensor network is established under the
following two application constraints:

\begin{itemize}[leftmargin=*] 
\item

\emph{spatial coverage bounds}; the number of deployed nodes at each
tier of the cluster-tree topology, $n$, is upper- and lower-bounded,
i.e. $N_{\textnormal{min}}\leq n\leq N_{\textnormal{max}}$

\item

\emph{temporal coverage lower bound}; the total frame acquisitions,
$k$, within a pre-defined time interval, $T$, is lower-bounded,
i.e. $k\geq K_{\text{min}}$

\end{itemize}

The bounds of the spatio--temporal coverage stem from application
specifics, such as: the cost of installing and maintaining visual
sensors, the minimum and maximum spatial coverage required for the
area to be monitored, and the minimum number of frames that allows
for visual data gathering and analysis with sufficient temporal resolution
within $T$ seconds.

Since the multimedia subsystem of each visual sensor produces varying
amounts of data depending on the monitored events and the specifics
of the visual analysis and processing under consideration, the bitstream
size produced by each sensor node in such multimedia applications
is a non-deterministic quantity. Therefore, the bitstream size produced
when each visual node processes $k$ frames within an activation interval
is a random variable (RV), $\mathcal{X}_{k}$, characterized by its
probability density function (PDF), $P(\chi_{k})$, $\mathcal{X}_{k}\backsim P(\chi_{k})$.
Since the underlying processes deriving this bitstream may not be
stationary and/or this data may include multi-rate channel codes (or
retransmissions) to alleviate channel impairments due to propagation
and other environmental effects of transmission, we assume marginal
statistics for $P\left(\chi_{k}\right)$, which are derived starting
from a doubly-stochastic model for the multimedia processing. Specifically,
such marginal statistics can be obtained by \cite{LamGoodmanDCT,foo2008analytical}:
\emph{(i)} fitting PDFs to sets of past measurements of bitstream
sizes transmitted by each sensor, with the statistical moments (parameters)
of such distributions characterized by another PDF; \emph{(ii)} integrating
over the parameter space to derive the final form of $P\left(\chi_{k}\right)$.
For example, if the bitstream size is modeled as a Half-Gaussian distribution
with variance parameter that is itself exponentially distributed,
by integrating over the parameter space, the marginal statistics of
the data rate become Laplacian \cite{LamGoodmanDCT,foo2008analytical}.

The disadvantage of using marginal statistics for the bitstream size
of each node during each activation interval is the removal of the
stochastic dependencies to its transient physical properties%
\footnote{e.g. the specifics of what is being monitored at each instant and
how the multimedia processing algorithm is operating on the input
data%
}. However, in this work we are interested in the \emph{expected} energy
consumption over a time interval and not in the instantaneous \emph{variations}
of energy consumption. Thus, a mean-based analysis using the marginal
statistics of the produced bitstream sizes is suitable for this purpose.

\subsection{Energy Consumption Penalties}

Following the push model of the camera node subsystem illustrated
in Figure \ref{fig:System_model}(b), each VSN node performs the following
operations:

\begin{enumerate}[leftmargin=*]
\item

\textit{Acquisition, processing and transmission: }A new frame is
acquired by means of a low-power camera sensor and processed with
a CPU-intensive algorithm, realized by the multimedia subsystem. Each
frame processing (possibly including coding to mitigate channel impairments)
produces, on average, $r$ bits for transmission. These bits are pushed
to the radio subsystem, which in turn transmits them to the higher
tier or, eventually, to the LPBR. Let $a$ Joule (J) be the energy
expenditure for acquiring a frame, $g$ be the average energy in Joule
(J) required for processing and producing one bit of information to
be transmitted and $j$ the average energy required to transmit it
to the LPBR or a relay node. Different multimedia applications may
incur different levels of energy consumption for the production of
each bit to be transmitted, while the average transmission energy
consumption per bit depends only on the specific radio chip used by
each wireless sensor node. Hence, the average energy consumed for
acquisition, processing and transmission within the active interval
of $T$ seconds is $ka+(g+j)\int_{0}^{\infty}\chi_{k}P(\chi_{k})d\chi=ka+(g+j)\mathbb{E}\left[\mathcal{X}_{k}\right]$
J, with $\mathbb{E}\left[\mathcal{X}_{k}\right]$ bits comprising
the statistical expectation of the data volume corresponding to $k$
frames.

\item

\textit{Buffering and Idling:} As shown in Figure \ref{fig:System_model},
each tier of the sensor network consists of $n$ sensor nodes that
communicate with the LPBR (or the relay nodes of the higher tier).
The set of all receivers (sink nodes) of each tier has predefined
consumption rate of $\frac{s}{T}$ kbps. Under balanced coupling,
each sensor node can transmit $\frac{s}{n}$ bits during the analysis
time interval of $T$ seconds. We thus identify two cases: if the
amount of data generated by the processing phase and relayed from
$d$ nodes of the lower tiers is less than $\frac{s}{n}$ bits, the
sensor node enters an ``idle'' state, where $b$ J/bit is consumed
for beaconing and other synchronization operations. The energy spent
during the idle mode of the analysis time interval is: $b\int_{0}^{\frac{s}{n}}(\frac{s}{n}-\chi_{k,d+1})P_{d+1}(\chi_{k,d+1})d\chi_{k,d+1}$
J, with $\mathcal{X}_{k,d+1}\sim P_{d+1}(\chi_{k,d+1})$ the RV modeling
the data rate of a node processing $k$ frames and relaying data from
$d$ other independent and identical nodes {[}with $\mathcal{X}_{k,1}\equiv\mathcal{X}_{k}$
and $P_{1}(\chi_{k})\equiv P(\chi_{k})${]}. Conversely, if the data
generated is greater than $\frac{s}{n}$ bits, then the sensor node
has to buffer the remaining data in a high-power, typically off-chip,
memory. Letting $p$ J be the energy cost of storing one bit of information,
the energy spent for buffering during the active time interval is:
$p\int_{\frac{s}{n}}^{\infty}(\chi_{k,d+1}-\frac{s}{n})P_{d+1}(\chi_{k,d+1})d\chi_{k,d+1}$
J. This case introduces delay, as buffered data will be scheduled
for later transmission. Thus, the proposed model is suitable for delay-tolerant
multimedia applications \cite{citeulike:3839709}.

\item

\emph{Receiving/Buffering and Relaying Data:} Under a multi-tier cluster-tree
topology, each node receives $d$ additional data streams from $d$
nodes positioned at the lower tier(s) and relays them along with its
own data streams (see Figure \ref{fig:System_model} for an example
with $d=2$). Over the analysis interval of $T$ seconds, the energy
expenditure corresponding to this process is given by $\left(h+j\right)\int_{0}^{\infty}\chi_{k,d}P_{d}\left(\chi_{k,d}\right)d\chi_{k,d}=\left(h+j\right)\mathbb{E}\left[\mathcal{X}_{k,d}\right]$
J, with $h$ J/bit the average energy required to receive and buffer
one bit and $\mathbb{E}\left[\mathcal{X}_{k,d}\right]$ the statistical
expectation of the number of bits received from all $d$ nodes of
the lower tier(s) during the active time interval. In practice, this
energy expenditure is dominated by the receiver power requirements%
\footnote{Energy rates $a$, $g$, $j$, $p$, $b$ and $h$ may also include
fixed, rate-independent costs of the particular multimedia or transceiver
hardware (e.g., visual sensor, transceiver or buffer startup and shutdown
costs).%
}. Given that, for IEEE 802.15.4-compliant transceivers, the transceiver
power under receive mode is virtually the same regardless if the node
is actually receiving data or not, it is irrelevant to the receiver
power whether the transmitting node(s) used their entire transmission
intervals or not.

\end{enumerate}

\begin{table}
\noindent \centering{}\caption{\label{tab:Nomenclature-table.1}Nomenclature table.}

\begin{tabular}{>{\centering}m{0.1\columnwidth}>{\centering}m{0.07\columnwidth}>{\raggedright}p{0.65\columnwidth}}
\multicolumn{1}{>{\centering}m{0.15\columnwidth}}{Symbol } & \multicolumn{1}{>{\centering}m{0.07\columnwidth}}{Unit} & \multicolumn{1}{>{\raggedright}m{0.6\columnwidth}}{Definition}\tabularnewline
\midrule 
$T$  & seconds  & Active time interval \tabularnewline
\midrule 
$n$, $N_{\min}$, $N_{\max}$  & --  & Number of transmitting sensor nodes \emph{at the same tier} of the
cluster-tree topology and minimum \& maximum nodes allowed by the
application \tabularnewline
\midrule 
$k$, $K_{\min}$  & --  & Number of frames captured and processed within $T$ seconds and minimum-allowed
by the application\tabularnewline
\midrule 
$r$  & bit  & Average number of bits produced after processing one frame\tabularnewline
\midrule 
$d$  & --  & Number of \emph{additional} nodes whose traffic is relayed by each
node at a given tier of the cluster-tree topology \tabularnewline
\midrule 
$a$  & J  & Energy to acquire one frame and initialize the multimedia processing\tabularnewline
\midrule 
$g$  & J/bit  & Energy for processing one bit\tabularnewline
\midrule 
$j$  & J/bit  & Energy for transmitting one bit\tabularnewline
\midrule 
$p$  & J/bit  & Penalty energy for storing one bit during receiver overloading\tabularnewline
\midrule 
$b$  & J/bit  & Energy during idle periods for the time interval corresponding to
one bit transmission\tabularnewline
\midrule 
$h$  & J/bit  & Energy for receiving and temporary buffering one bit under the relay
case\tabularnewline
\midrule 
$s$  & bit  & Data volume (bits) of a relay node (or base station) received within
$T$ seconds\tabularnewline
\midrule 
$\mathcal{X}_{k,d+1}\sim$

$P_{d+1}\left(\chi_{k,d+1}\right)$  & bit  & RV modeling the cumulative bits transmitted by each node, including
the bits relayed from $d$ nodes of lower tiers, after each node processed
$k$ video frames \tabularnewline
\midrule 
$\mathbb{E}\left[\mathcal{X}_{k,d+1}\right]$  & bit  & Statistical expectation of $\mathcal{X}_{k,d+1}$\tabularnewline
\midrule 
$E_{c}$  & J  &  Energy consumption of \textsl{each individual node} over the analysis
time interval $T$\tabularnewline
\midrule 
$\beta_{\text{D}}$, $\gamma_{\text{D}}$  & --  & Parameters expressing the combination of the system energy rates,
receiver rate and the mean of the utilized marginal PDF $\text{D}$
for the solutions obtained along the spatial and temporal direction \tabularnewline
\bottomrule
\end{tabular}
\end{table}

\subsection{Expected Energy Consumption}

Summing all contributions 1\textasciitilde{}3 of the previous subsection,
the energy consumption of each node, $E_{\text{c}}$, over the time
interval $T$ is:

\begin{align}
E_{\text{c}} & \left(n,\, k\right)=ka+(g+j)\mathbb{E}[\mathcal{X}_{k}]+(h+j)\mathbb{E}[\mathcal{X}_{k,d}]\nonumber \\
 & +p\int_{\frac{s}{n}}^{\infty}(\chi_{k,d+1}-\frac{s}{n})P_{d+1}(\chi_{k,d+1})d\chi_{k,d+1}\label{eq:e_c1}\\
 & +b\int_{0}^{\frac{s}{n}}(\frac{s}{n}-\chi_{k,d+1})P_{d+1}(\chi_{k,d+1})d\chi_{k,d+1}.\nonumber 
\end{align}
Adding and subtracting $p\int_{0}^{\frac{s}{n}}(\chi_{k,d+1}-\frac{s}{n})P_{d+1}(\chi_{k,d+1})d\chi_{k,d+1}$
to \eqref{eq:e_c1} leads to:

\begin{align}
E_{\text{c}}\left(n,\, k\right) & =ka+(g+j)\mathbb{E}[\mathcal{X}_{k}]+(h+j)\mathbb{E}[\mathcal{X}_{k,d}]\nonumber \\
 & +p\mathbb{E}[\mathcal{X}_{k,d+1}]-\frac{ps}{n}\label{eq:e_c2}\\
 & +(b+p)\int_{0}^{\frac{s}{n}}(\frac{s}{n}-\chi_{k,d+1})P_{d+1}(\chi_{k,d+1})d\chi_{k,d+1}.\nonumber 
\end{align}
Since the VSN is uniformly formed, all sensors are independent and
identical. We can thus establish the relationships:

\begin{equation}
\forall d>0:\:\mathbb{E}\left[\mathcal{X}_{k,d+1}\right]=\frac{d+1}{d}\mathbb{E}\left[\mathcal{X}_{k,d}\right],\label{eq:E_d+1 vs E_d}
\end{equation}

\begin{equation}
\forall d>0:\:\mathbb{E}\left[\mathcal{X}_{k,d}\right]=d\mathbb{E}\left[\mathcal{X}_{k}\right],\label{eq:E_d+1 vs E_d-1}
\end{equation}
which are based on the fact that the expected number of bits transmitted
or received by a node increases linearly with respect to $d$. By
modifying \eqref{eq:e_c2} based on \eqref{eq:E_d+1 vs E_d} and \eqref{eq:E_d+1 vs E_d-1},
we reach:

\begin{align}
E_{\text{c}}\left(n,\, k\right) & =ka+\left[(p+j)\left(d+1\right)+hd+g\right]\mathbb{E}[\mathcal{X}_{k}]-\frac{ps}{n}\nonumber \\
 & +(b+p)\int_{0}^{\frac{s}{n}}(\frac{s}{n}-\chi_{k,d+1})P_{d+1}(\chi_{k,d+1})d\chi_{k,d+1}.\label{eq:e_c2-1}
\end{align}
This equation is the basis for the analytic exploration of the minimum
energy consumption under several marginal PDFs characterizing the
data production and transmission process.

\section{Analytic Derivation of Minimum Energy Consumption}

\label{sec:min_energy_analysis}

Our objective is to derive the spatio--temporal parameters minimizing
$E_{\text{c}}\left(n,k\right)$ in \eqref{eq:e_c2-1}, subject to
the spatio--temporal constraints defined in Section \ref{sec:System Model},
that is:


\begin{align}
\left\{ n^{\star},k^{\star}\right\}  & =\arg\underset{\forall n,k}{\text{min}}E_{\text{c}}\left(n,k\right),\label{eq:arg_min_Ec}
\end{align}
with

\begin{equation}
N_{\textnormal{min}}\le n\le N_{\textnormal{max}}\;\text{and }k\ge K_{\text{min}}\label{eq:n_k_constraints}
\end{equation}
and $\left\{ n^{\star},k^{\star}\right\} $ the values deriving the
minimum energy consumption.

In the following, we consider different distributions for $P_{d+1}\left(\chi_{k,d+1}\right)$
and derive the solution for $n$ and $k$ that minimizes the energy
consumption, while ensuring the conditions imposed by the application
constraints are met. While our analysis is assuming that $n$ and
$k$ are continuous variables, once the $\left\{ n^{\star},k^{\star}\right\} $
values are derived, they can be discretized to the points $\left\{ \left\lfloor n^{\star}\right\rfloor ,\left\lfloor k^{\star}\right\rfloor \right\} $,
$\left\{ \left\lceil n^{\star}\right\rceil ,\left\lceil k^{\star}\right\rceil \right\} $
$\left\{ \left\lceil n^{\star}\right\rceil ,\left\lfloor k^{\star}\right\rfloor \right\} $
$\left\{ \left\lfloor n^{\star}\right\rfloor ,\left\lceil k^{\star}\right\rceil \right\} $
{[}if all four satisfy the constraints of \eqref{eq:n_k_constraints}{]}
in order to check which discrete pair of values derives the minimum
energy consumption in \eqref{eq:e_c2-1}. This is because: \emph{(i)}
the energy functions under consideration are continuous and differentiable;
and \emph{(ii)} we shall show that a unique minimum is found for \eqref{eq:e_c2-1}
that is parametric to the setting of the temporal constraint ($K_{\text{min}}$).
As such, the analysis on the continuous variable space can be directly
mapped onto the discrete variable set under the aforementioned discretization.

\subsection{Definitions of Data Transmission PDFs under Consideration and Infeasibility
of Global Minimum of $E_{\text{c}}\left(n,k\right)$ }

When one has limited or no knowledge about the cumulative data transmitted
by each VSN node during the active time interval, one can assume that
$P_{d+1}(\chi_{k,d+1})$ is uniform over the interval $[0,\,2kr\left(d+1\right)]$. 
\begin{defn}
\emph{($P_{d+1}(\chi_{k,d+1})$ is Uniform):} \emph{We define $P_{d+1}(\chi_{k,d+1})$
as the Uniform distribution when:} 
\begin{equation}
P_{d+1}(\chi_{k,d+1})=\begin{cases}
\frac{1}{2kr\left(d+1\right)} & \,\,\,0\le\chi_{k,d+1}\le2kr\left(d+1\right)\\
0 & \,\,\,\text{otherwise}
\end{cases}\label{eq:uniform}
\end{equation}
\emph{with }$\E_{\text{U}}[\mathcal{X}_{k}]=kr$ \emph{{[}and }$\E_{\text{U}}[\mathcal{X}_{k,d+1}]=kr\left(d+1\right)$\emph{{]}
corresponding to the mean value of the data transmitted by a node
that produces $k$ frames of $r$ bits each on average (and relays
information from $d$ other nodes).} 
\end{defn}
\textbf{Corollary 1.}\emph{ When }$P_{d+1}(\chi_{k,d+1})$ \emph{is
Uniform, there exists no global solution to \eqref{eq:arg_min_Ec}
in its unconstrained form.} 
\begin{IEEEproof}
Using \eqref{eq:uniform} in \eqref{eq:e_c2-1} leads to: 
\begin{align}
E_{\text{c,U}}(n,k) & =k\left[a+r\left[(p+j)(d+1)+hd+g\right]\right]\label{eq:uniform_energy}\\
 & -\frac{ps}{n}+\frac{s^{2}(b+p)}{4n^{2}kr(d+1)}.\nonumber 
\end{align}

To obtain the solution to \eqref{eq:arg_min_Ec} under the energy
consumption given by \eqref{eq:uniform_energy}, one can search for
critical points of $E_{\text{c,U}}$. By definition, a critical point
of a multidimensional function is the point where the gradient of
the function is equal to zero. Imposing that the derivatives of $E_{\text{c,U}}$
with respect to $n$ and $k$ are both equal to zero leads to:

\begin{equation}
\left\{ \begin{array}{l}
\frac{\partial E_{\text{c,U}}}{\partial n}=\frac{ps}{n^{2}}-\frac{s^{2}(b+p)}{2n^{3}kr(d+1)}=0\\
\frac{\partial E_{\text{c,U}}}{\partial k}=a+r\left[(p+j)(d+1)+hd+g\right]\\
\qquad\;-\frac{s^{2}(b+p)}{4n^{2}k^{2}r(d+1)}=0
\end{array}\right.\label{eq:gradient_zero}
\end{equation}

Solving $\frac{\partial E_{\text{c,U}}}{\partial n}=0$ for $n$ gives
$n=\frac{s\left(b+p\right)}{2krp\left(d+1\right)}$. Substituting
this solution in $\frac{\partial E_{\text{c,U}}}{\partial k}=0$ and
solving for $a$, leads to $a<0$. However, this is not feasible since
$a$ is the energy cost to acquire one frame. Hence, under the physical
constraints of the problem, \emph{there is no single (global) solution
$\left\{ n^{\star},k^{\star}\right\} \in\mathcal{\mathbb{R\times R}}$
to \eqref{eq:arg_min_Ec} in its unconstrained form}, i.e. when one
ignores the constraints of \eqref{eq:n_k_constraints}. 
\end{IEEEproof}
We now extend the analysis towards other PDFs for the data transmission,
which are frequently encountered in practice. 
\begin{defn}
\emph{($P_{d+1}(\chi_{k,d+1})$ is Pareto):} \emph{We consider $P_{d+1}(\chi_{k,d+1})$
as the Pareto distribution with scale $v$ and shape $\alpha>1$ when:
\begin{equation}
P_{d+1}(\chi_{k,d+1})=\begin{cases}
\alpha\frac{v^{\alpha}}{\chi_{k,d+1}^{\alpha+1}}, & \,\,\chi_{k,d+1}\ge v\\
0, & \,\,\text{otherwise}
\end{cases}.\label{eq:Pareto}
\end{equation}
Setting $v=\frac{\alpha-1}{\alpha}kr(d+1)$ leads to} $\E_{\text{P}}[\mathcal{X}_{k}]=kr$\emph{,
i.e. we match the expected data volume to that of the Uniform PDF.} 
\end{defn}
The Pareto distribution has been used, amongst others, to model the
marginal data size distribution of TCP sessions that contain substantial
number of small files and a few very large ones \cite{Paxson95wide-areatraffic:},
\cite{Park96onthe}. It has also been used to model multimedia traffic
packet sizes in several works, e.g. by Kumar \cite{Kumar:2003:PTS:1170745.1171535}. 
\begin{defn}
\emph{(}$P_{d+1}(\chi_{k,d+1})$ is \emph{Exponential):} \emph{We
consider $P_{d+1}(\chi_{k,d+1})$ as the Exponential distribution}
\emph{when:}

\emph{
\begin{equation}
P_{d+1}(\chi_{k,d+1})=\frac{1}{kr(d+1)}\exp\left(-\frac{1}{kr(d+1)}\chi_{k,d+1}\right).\label{eq:Exponential PDF definition}
\end{equation}
with }$\E_{\text{E}}[\mathcal{X}_{k}]=kr$ \emph{{[}and }$\E_{\text{E}}[\mathcal{X}_{k,d+1}]=kr\left(d+1\right)$\emph{{]}
corresponding to the mean value of the data transmitted by a node
that produces $k$ frames of $r$ bits each on average (and relays
information from $d$ other nodes).} 
\end{defn}
We remark that the marginal statistics of MPEG video traffic have
often been modeled as exponentially decaying \cite{Dai:2009:UTM:1653034.1653052}.

We conclude by considering $P_{d+1}(\chi_{k,d+1})$ as the Half-Gaussian
distribution with mean $\E_{\text{H}}[\mathcal{X}_{k}]=kr$. This
distribution has been widely used in data gathering problems in science
and engineering when the modeled data has non-negativity constraints.
Some recent examples include the statistical characterization of motion
vector data sizes in Wyner-Ziv video coding algorithms suitable for
VSNs \cite{DBLP:conf/icip/TagliasacchiTS06}, or the statistical characterization
of sample amplitudes captured by an image sensor \cite{LamGoodmanDCT,foo2008analytical,DBLP:journals/tip/AndreopoulosP08}. 
\begin{defn}
\emph{($P_{d+1}(\chi_{k,d+1})$ is Half-Gaussian):} \emph{We consider
$P_{d+1}(\chi_{k,d+1})$ as the Half-Gaussian distribution when:} 
\end{defn}
\emph{
\begin{equation}
P_{d+1}(\chi_{k,d+1})=\begin{cases}
\frac{2}{\pi kr(d+1)}\exp\left(-\frac{\chi_{k,d+1}^{2}}{\pi k^{2}r^{2}(d+1)^{2}}\right), & \chi_{k,d+1}\geq0\\
0, & \chi_{k,d+1}<0
\end{cases}\label{eq:halfgaussain PDF definition}
\end{equation}
with }$\E_{\text{H}}[\mathcal{X}_{k}]=kr$ \emph{{[}and }$\E_{\text{H}}[\mathcal{X}_{k,d+1}]=kr\left(d+1\right)$\emph{{]}
corresponding to the mean value of the data transmitted by a node
that produces $k$ frames of $r$ bits each on average (and relays
information from $d$ other nodes).}

\textbf{Corollary 2.}\emph{ When }$P_{d+1}(\chi_{k,d+1})$ \emph{is
the Pareto, Exponential or Half-Gaussian distribution, given by \eqref{eq:Pareto}--\eqref{eq:halfgaussain PDF definition},
there exists no global solution to \eqref{eq:arg_min_Ec} in its unconstrained
form.} 
\begin{IEEEproof}
Under \eqref{eq:Pareto}, the energy expression of \eqref{eq:e_c2-1}
becomes: 
\begin{align}
E_{\text{c,P}} & =k\left[a+r\left[\left(p+j\right)\left(d+1\right)+hd+g\right]\right]\nonumber \\
 & +\frac{bs}{n}+\left(b+p\right)\left(\frac{v^{\alpha}n^{\alpha-1}}{s^{\alpha-1}\left(\alpha-1\right)}-\frac{\alpha v}{\alpha-1}\right).\label{eq:pareto_energy}
\end{align}

In addition, replacing \eqref{eq:Exponential PDF definition} in the
energy expression of \eqref{eq:e_c2-1}, we obtain: 
\begin{align}
E_{\text{c,E}} & =k\left[a+r\left[\left(p+j\right)\left(d+1\right)+hd+g\right]\right]+\frac{bs}{n}\label{eq:exponential_energy}\\
 & +\left(b+p\right)\left[kr\left(d+1\right)\left(\exp\left(-\frac{s}{nkr(d+1)}\right)-1\right)\right].\nonumber 
\end{align}

Finally, replacing \eqref{eq:halfgaussain PDF definition} in the
energy expression of \eqref{eq:e_c2-1}, we obtain:

\begin{equation}
\begin{array}{cc}
E_{\text{c,H}}=k\left[a+\left[r\left(p+j\right)\left(d+1\right)+hd+g\right]\right]-\frac{ps}{n}+\left(b+p\right)\\
\times\left[kr\left(d+1\right)\left(\exp\left(-\frac{s^{2}}{\pi k^{2}r^{2}n^{2}(d+1)^{2}}\right)-1\right)\right.\\
\left.+\frac{s}{n}\text{erf}\left(\frac{s}{\sqrt{\pi}krn(d+1)}\right)\right].
\end{array}\label{eq:halfgaussian_energy}
\end{equation}

To obtain the solution to \eqref{eq:arg_min_Ec} under the energy
consumption given by \eqref{eq:uniform_energy}, one can search for
critical points of $E_{\text{c,P}}$, $E_{\text{c,E}}$ and $E_{\text{c,H}}$.
Similarly as for Corollary 1, it is straightforward to show that imposing
that the derivatives of $E_{\text{c,P}}$, $E_{\text{c,E}}$ and $E_{\text{c,H}}$
with respect to $n$ and $k$ are both equal to zero leads to solutions
that require $a<0$ (detailed derivations omitted), which is not physically
feasible since $a$ is the energy cost to acquire one frame. 
\end{IEEEproof}
It follows from Corollary 1 and 2 that, under the physical constraints
of the problem, \emph{there is no single (global) solution $\left\{ n^{\star},k^{\star}\right\} \in\mathcal{\mathbb{R\times R}}$
to \eqref{eq:arg_min_Ec} in its unconstrained form}, i.e., when one
ignores the constraints of \eqref{eq:n_k_constraints}. However, we
may consider each dimension individually (i.e., perform univariate
minimization along the $n$ or $k$ dimension) in order to find a
local or global minimum for that particular dimension and then choose
for the other dimension the value that minimizes \eqref{eq:arg_min_Ec}
under the spatio--temporal constraints of \eqref{eq:n_k_constraints}.
Subsequently, we can identify if the derived minima are unique under
the imposed constraints and whether the entire region of support of
the energy function under these constraints has been covered by the
derived solutions. Following this approach, the main results are presented
in the following subsection. The detailed derivations are contained
in the Appendices.

\subsection{Main Results: Parametric Minima of $E_{\text{c}}\left(n,k\right)$ }

\textbf{Proposition 1.}\emph{ When the data transmitted by each VSN
node follows the Uniform, Pareto or Exponential distributions of Definitions
1--3, the sets of solutions giving the minimum energy consumption
in \eqref{eq:arg_min_Ec} under the spatio--temporal constraints of
\eqref{eq:n_k_constraints} are:} 
\begin{equation}
\left\{ n^{\star},k^{\star}\right\} _{\text{D}}=\begin{cases}
\left(N_{\text{max}},\,\frac{\gamma_{\textrm{D}}}{N_{\text{max}}}\right) & \text{if}\,\, K_{\text{min}}\le\frac{\gamma_{\textrm{D}}}{N_{\text{max}}}\\
\left(N_{\text{max}},\, K_{\text{min}}\right) & \text{if}\,\,\frac{\gamma_{\textrm{D}}}{N_{\text{max}}}<K_{\text{min}}<\frac{\beta_{\text{D}}}{N_{\max}}\\
\left(\frac{\beta_{\text{D}}}{K_{\min}},\, K_{\text{min}}\right) & \text{if}\,\,\frac{\beta_{\text{D}}}{N_{\max}}\le K_{\text{min}}\le\frac{\beta_{\text{D}}}{N_{\min}}\\
\left(N_{\text{min}},\, K_{\text{min}}\right) & \text{if}\,\, K_{\text{min}}>\frac{\beta_{\text{D}}}{N_{\min}}
\end{cases}\label{eq:uniform_optimal-1}
\end{equation}
\emph{with }$\text{D}\in\left\{ \text{U},\,\text{P},\,\text{E}\right\} $
\emph{indicating each of the three distributions, and }$\beta_{\text{D}}$
\emph{and }$\gamma_{\text{D}}$ \emph{defined by:} 
\begin{equation}
\beta_{\text{U}}=\frac{s\left(b+p\right)}{2pr\left(d+1\right)},\label{eq:beta_U}
\end{equation}
\begin{equation}
\gamma_{\textrm{U}}=\frac{s}{2}\sqrt{\frac{b+p}{r\left(d+1\right)\left[a+r\left[\left(p+j\right)\left(d+1\right)+hd+g\right]\right]}},\label{gamma_U}
\end{equation}
\begin{equation}
\gamma_{\textrm{P}}=\frac{s\alpha}{r(\alpha-1)(d+1)}\left(\frac{r\left[\left(b-j\right)\left(d+1\right)-hd-g\right]-a}{r\left(d+1\right)\left(b+p\right)}\right)^{\frac{1}{\alpha-1}},\label{eq:gamma_P}
\end{equation}
\begin{equation}
\beta_{\text{P}}=\frac{s\alpha}{r\left(\alpha-1\right)\left(d+1\right)}\left(\frac{b}{b+p}\right){}^{\frac{1}{\alpha}},\label{eq:beta_P}
\end{equation}

\begin{equation}
\beta_{\text{E}}=\frac{s}{r(d+1)\ln(\frac{b+p}{p})}\label{eq:beta_E}
\end{equation}
\emph{and}

\begin{equation}
\gamma_{\textrm{E}}=-\frac{s}{r(d+1)\left[W\left(-\frac{1}{\exp}\frac{-a+r\left[\left(b-j\right)\left(d+1\right)-hd-g\right]}{r(d+1)(b+p)}\right)+1\right]},\label{eq:gamma_E}
\end{equation}
\emph{with $W(\cdot)$ the Lambert product-log function \cite{corless1996lambertw}.
For the particular case when }$\text{D}=\text{E}$ \emph{(Exponential
PDF), \eqref{eq:uniform_optimal-1} holds under the condition that
$p>b$, i.e., the penalty energy to buffer bits is higher than beaconing
energy.} 
\begin{IEEEproof}
See Appendix \ref{sec:Appendix-I}. 
\end{IEEEproof}
\textbf{Proposition 2.}\emph{ When the data transmitted by each VSN
node follows the Half-Gaussian distribution of Definition 4, the set
of solutions giving the minimum energy consumption in \eqref{eq:arg_min_Ec}
under the spatio--temporal constraints of \eqref{eq:n_k_constraints}
is:}

\begin{equation}
\left\{ n^{\star},k^{\star}\right\} _{\text{H}}=\begin{cases}
\left(N_{\text{max}},\, K_{\text{min}}\right) & \text{if}\,\, K_{\text{min}}\le\frac{\beta_{\text{H}}}{N_{\max}}\\
\left(\frac{\beta_{\text{H}}}{K_{\text{min}}},\, K_{\text{min}}\right) & \text{if}\,\,\frac{\beta_{\text{H}}}{N_{\max}}<K_{\text{min}}\leq\frac{\beta_{\text{H}}}{N_{\text{min}}}\\
\left(N_{\textrm{min}},\, K_{\text{min}}\right) & \text{if}\,\, K_{\text{min}}>\frac{\beta_{\text{H}}}{N_{\text{min}}},
\end{cases}\label{eq:halfgaussian_optimal}
\end{equation}
\emph{with} 
\begin{equation}
\beta_{\text{H}}=\frac{s}{\sqrt{\pi}r\left(d+1\right)\text{ erf}^{-1}\left(\frac{p}{b+p}\right)}.\label{eq:beta_H}
\end{equation}

\begin{IEEEproof}
The proof follows the same steps as for the previous cases and it
is summarized in Appendix \ref{sub:Appendix-Half-Gaussian}. 
\end{IEEEproof}

\subsection{Discussion}

The key observation from Propositions 1 and 2 is that, regardless
to the distribution used for modeling the data production process,
the solutions giving the minimum energy consumption attain the same
mathematical form. Specifically, when the initial constraint on the
minimum number of frames captured and processed within $T$ seconds,
$K_{\min}$, is higher than the threshold value: $\frac{\beta_{\text{D}}}{N_{\min}}$,
the optimal solution is the one where $N_{\min}$ nodes process $K_{\min}$
frames each (i.e., the minimum setting possible for nodes and frames-per-node).
If $K_{\min}$ is smaller or equal than this threshold, therefore
facilitating more nodes within each tier of the VSN, the optimal number
of nodes, $n^{\star}$, derived by Propositions 1 and 2, increases
to $\frac{\beta_{\text{D}}}{K_{\min}}$. However, when $n^{\star}$
reaches the constraint on the maximum number of nodes, $N_{\max}$,
then the optimal solution for each node is to use a frame setting
that is higher than $K_{\min}$. The latter is true for Proposition
1; however, for Proposition 2 (Half-Gaussian PDF), the corresponding
optimal frame setting was found to be imaginary regardless to the
specific system parameter. Therefore, the optimal solution for this
case is always $k^{\star}=K_{\min}$.

In terms of relevance to practical applications, the results of this
section can be used to assess the impact of the spatio--temporal constraints
and the data production and transmission process (as characterized
by its marginal PDF) on the energy consumption of VSNs, under a variety
of energy consumption rates for the radio and multimedia subsystems.
For example, under given energy availability from the node battery
and predetermined system activation time ($T$), this allows for the
determination of appropriate hardware to be used (i.e. $j$, $h$,
$b$, $p$, $a$ and $g$ parameters) in order to meet the spatio--temporal
constraints of the application. Moreover, via the analysis of the
previous four subsections, one can optimize the system under the assumption
of a certain marginal PDF characterizing the data production and transmission
process of each node.

Conversely, under particular technology (i.e. given $j$, $h$, $b$,
$p$, $a$ and $g$ parameters) and given configuration for the VSN
in terms of number of nodes and frames to capture within the activation
time interval, one can determine the required energy in order to achieve
the designated visual data gathering task. Furthermore, under the
proposed framework, one can determine the data production and transmission
(marginal) PDFs that meet predetermined energy supply and spatio--temporal
constraints. 

Although we do not claim that the utilized PDFs cover all possible
scenarios that can be encountered in practice, they comprise an ensemble
of distributions that includes several important aspects, i.e.,: \emph{(i)}
the maximum-entropy PDF (Uniform); \emph{(ii)} well-known distributions
characterizing the transmission rate of real-world systems (Exponential
and Half-Gaussian) \cite{LamGoodmanDCT,foo2008analytical,DBLP:journals/tip/AndreopoulosP08,DBLP:conf/icip/TagliasacchiTS06,Dai:2009:UTM:1653034.1653052},
and \emph{(iii)} a parameterized distribution (Pareto) that corresponds
to the continuous equivalent to Zipf's law for generalized frequency
of occurrences of physical phenomena; moreover, if $\alpha=kr$, the
Pareto distribution corresponds to near fixed-rate transmission with
rate $kr$. Beyond the cases considered in this paper, if another
distribution provides a better fit to a particular deployment, the
steps of Propositions 1 and 2 can be used to provide a characterization
of the available solution space. Moreover, given that the results
of Propositions 1 and 2 are applicable per node, if the considered
scenario involves a non uniformly-formed VSN, the same analysis applies
for each node of each cluster-tree tier, albeit with the use of: 
\begin{enumerate}
\item a different PDF per sensor, leading to a mixture of PDFs for the relayed
traffic, with the resulting distribution being the convolution of
the intermediate distributions; 
\item \emph{unbalanced} \emph{coupling} in \eqref{eq:e_c2} and \eqref{eq:E_d+1 vs E_d},
i.e., the $i$th node transmitting $s_{i}$ bits during the analysis
time interval of $T$ seconds, with $s_{i}$ allocated by the utilized
protocol during the cluster formation \cite{AlvesClusterTreeGTS,citeulike:3839709,koubaa2006gts,LeungMACEssentials}; 
\item the $i$th node of each cluster relaying traffic from $d_{i}$ nodes,
and, in general, $d_{i}\neq d_{i^{\shortmid}}$ for $i\neq i^{\shortmid}$. 
\end{enumerate}
Given that a numerical package (e.g., Mathematica or Matlab Symbolic)
can be used for the calculation of: \emph{(i)} the convolution of
$d_{i}+1$ distributions $P_{d_{i}+1}\left(\chi_{k,d_{i}+1}\right)$
(corresponding to the mixture of $d_{i}+1$ PDFs of the $i$th node
of each tier) and \emph{(ii)} the $\int_{0}^{s_{i}}\left(s_{i}-\chi_{k,d_{i}+1}\right)P_{d_{i}+1}\left(\chi_{k,d_{i}+1}\right)d\chi_{k,d_{i}+1}$
term of \eqref{eq:e_c2}, we do not expand on these cases further.

Overall, our proposed energy consumption model and the associated
analytic results can be used in many ways for early-stage exploration
of system, network, and data production parameters in VSNs that match
the design specifications of classes of application domains. Such
application examples are given in Section \ref{sec:Applications}.

\section{Evaluation of the Analytic Results\label{sec:Evaluation-of-Energy} }

To validate the proposed analytic model of \eqref{eq:e_c2-1} and
Propositions 1 and 2 for the settings leading to the minimum energy
consumption, we performed a series of experiments based on a visual
sensor network matching the system model of Section \ref{sec:System Model}
and an energy-measurement testbed. Specifically, each visual node
of the sensor network is composed of a BeagleBone Linux Computer (multimedia
subsystem) attached to a TelosB sensor node for low-power wireless
communications (radio subsystem) \cite{CancliniSENSYS2013}. Each
BeagleBone is equipped with a RadiumBoard CameraCape to provide for
the video frame acquisition. For energy-efficient processing, we downsampled
all input images to QVGA (320x240) resolution. 

In order to measure the energy consumption of each VSN node, we captured
the real-time current consumption at two high-tolerance 1 Ohm resistors,
the first of which was placed in series with the multimedia and the
second in series with the radio subsystem of each visual node. A Tektronix
MDO4104-6 oscilloscope was used for the two current consumption captures
of each experiment. Further, our deployment involved: \emph{(i)} a
TelosB node serving as the LPBR and collecting all bitstreams and
2 to 32 visual nodes positioned within four adjacent rooms and the
corridor of the same floor of the Department of Electrical and Electronic
Engineering at University College London {[}following the layout of
Figure \ref{fig:System_model}(a){]}; \emph{(ii)} a uniformly-formed
hierarchical cluster-tree network topology with $n=2$ to $n=16$
nodes per network tier and the recently-proposed (and available as
open source) TFDMA protocol \cite{burana2012DTFDMA} for contention-free
MAC-layer coordination; \emph{(iii)} no WiFi or other IEEE802.15.4
networks concurrently operating in the utilized channels of the 2.4
GHz band. Even if IEEE802.11 or other IEEE802.15.4 networks coexist
with the proposed deployment, well-known channel hopping schemes like
TSCH \cite{6185525} or interference-avoidance schemes \cite{5672592}
can be used at the MAC layer to mitigate such external interference
while maintaining a balanced cluster tree topology in the WSN. 

TFDMA ensures collision-free multichannel communications with guaranteed
timeslots via a fair time-division multiple access (TDMA) schedule
constructed within each of the utilized channels of the IEEE802.15.4
physical layer via beacon packet exchanges \cite{burana2012DTFDMA}.
Protocols such as TFDMA, the TSCH mode of IEEE 802.15.4e-2012 \cite{tinka2010TSCH}
and other balanced cluster-tree--based MAC-layer protocols \cite{AlvesClusterTreeGTS,wang2012IPV6,zuo2012two,koubaa2006gts},
allow for collision-free, uniformly-formed, cluster-tree based VSNs
to be formed via the combination of fair TDMA scheduling and channel
allocation or channel hopping. Experiments have shown that such protocols
can scale to hundreds or even thousands of nodes \cite{pister2008tsmp}.
Therefore, our evaluation is pertinent to such scenarios that may
be deployed in the next few years within the IoT paradigm \cite{zorzi2010IoT,gubbi2013IoT}.

\subsection{Radio Subsystem }

For what concerns the radio subsystem, each TelosB runs the low-power
Contiki 2.6 operating system. Given that the utilized TFDMA protocol
ensures collision-free transmissions from each node, we enabled the
low-power NullMAC and NullRDC options of the Contiki OS that disable
the default MAC queuing and backoff mechanisms. This led to data consumption
rate at the application layer of $\frac{s}{T}=144$ kbps. 

Given that varying the transmission power level has minimal effect
on the VSN node energy consumption (since most of the transceiver
current consumption is due to reception) and may compromise error-free
data reception, we utilized the maximum transmit power, which led
to reliable data transmission under the collision-free timeslot allocation
of TFDMA. Under these operational settings, the average transmission
cost per bit of information, $j$ J/bit, as well as the cost for beaconing,
$b$ J/bit, and buffering, $p$ J/bit, were established experimentally
by repeating several dedicated energy-measurement tests with the TelosB
subsystem; their values are shown at the top half of Table \ref{tab:TelosB Settings}
and we have experimentally verified that they remained constant over
several activation intervals.

\subsection{Multimedia Subsystem }

Since the energy consumption of the multimedia subsystem is application-dependent,
we focused on two different applications, namely: \textit{(i)} encoding
and transmission of JPEG video frames and \textit{(ii)} extraction
and transmission of local features for visual analysis. These two
scenarios represent a wide range of practical VSN-related deployments
proposed recently \cite{charfi2009VSN,kwon2009imagesensor,mingorance2010efficient,paniga2011experimental,rahimi2005cyclops,redondi2012rate,yu2004energy,DBLP:conf/MMSP/RedondiBCT}. 

\begin{enumerate}[leftmargin=*] 
\item

\textit{Differential Motion JPEG (MJPEG) encoding:} We used a hybrid
DCT-DPCM encoder, such as the one presented in \cite{paniga2011experimental}.
In this system, the first frame of the video sequence is JPEG encoded
and transmitted. For the subsequent frames, only the difference between
two adjacent frames is encoded. The encoding process follows the standard
JPEG baseline, i.e., quantization of the Discrete Cosine Transform
(DCT) coefficients followed by run length coding (RLE) and Huffman
coding. 

\item

\textit{Visual Features extraction: }Several visual analysis tasks
can be performed by disregarding the pixel representation of an image,
and relying only on a much more compact representation based on local
visual features \cite{DBLP:journals/ijcv/Lowe04}. In a nutshell,
salient keypoints of an image are identified by means of a detector,
and a descriptor is computed from the pixel values belonging to the
image patch around each keypoint. Here, we focus on corner-like local
features produced by processing each frame of the input video sequence
with the FAST corner detector \cite{rosten_2008_faster}, which is
optimized for fast extraction of visual features on low-power devices.
Each detected keypoint is then described by means of a binary descriptor:
we used the BRIEF algorithm \cite{calonder2010brief}, which outputs
descriptors of 64 bytes each. 

\end{enumerate}

Dedicated energy-measurement tests were performed with the Beaglebone
multimedia subsystem by varying the encoding quality factor for differential
MJPEG, while for features extraction, we varied the FAST detection
threshold. This allowed us to trace curves in the energy-rate plane
and to obtain the average energy cost per bit, as well as the average
initialization cost per frame for both the application scenarios,
which are reported at the bottom half of Table \ref{tab:TelosB Settings}.
The cost of acquiring one frame was derived from the specifications
of the AptinaMT9M114 image sensor mounted on the CameraCape and is
reported in Table \ref{tab:TelosB Settings}. The overall acquisition
cost for one frame is established as $a=a_{\textrm{ACQ}}+a_{\textrm{JPEG}}$
for the JPEG case and $a=a_{\textrm{ACQ}}+a_{\textrm{VF}}$ for the
visual-feature extraction case. 

\begin{table}[t!]
\begin{centering}
\caption{Visual sensor energy and bitrate parameters.\label{tab:TelosB Settings}}

\par\end{centering}

\centering{}%
\begin{tabular}{|c|c|c|c|}
\hline 
\textbf{Parameter}  & \textbf{Description}  & \textbf{Unit}  & \textbf{Value }\tabularnewline
\hline 
\hline 
\multicolumn{4}{|c|}{\textbf{Radio Subsystem (TelosB)}}\tabularnewline
\hline 
\hline 
$\frac{s}{T}$  & Data consumption rate  & kbps  & 144\tabularnewline
\hline 
$j$  & Transmission cost  & J/bit  & $2.20\times10^{-7}$\tabularnewline
\hline 
$h$  & Receiving cost  & J/bit  & $2.92\times10^{-6}$\tabularnewline
\hline 
$b$  & Beaconing/idling cost  & J/bit  & $1.90\times10^{-7}$\tabularnewline
\hline 
$p$  & Buffering cost  & J/bit  & $2.86\times10^{-7}$\tabularnewline
\hline 
\hline 
\multicolumn{4}{|c|}{\textbf{Multimedia Subsystem (BeagleBone)}}\tabularnewline
\hline 
\hline 
$a_{\textrm{ACQ}}$  &  Acquisition cost  & J  & $5.00\times10^{-3}$\tabularnewline
\hline 
$a_{\textrm{JPEG}}$  & Initialization cost (JPEG)  & J  & $1.40\times10^{-2}$\tabularnewline
\hline 
$a_{\textrm{VF}}$  & Initialization cost (Visual Feat.) & J & $7.79\times10^{-3}$\tabularnewline
\hline 
$g_{\textrm{JPEG}}$  & Processing cost (JPEG)  & J/bit  & $4.40\times10^{-8}$\tabularnewline
\hline 
$g_{\textrm{VF}}$  & Processing cost (Visual Feat.)  & J/bit  & $1.90\times10^{-8}$\tabularnewline
\hline 
\end{tabular}
\end{table}

\begin{figure*}[t]
\centering \subfigure[Uniform PDF ($d = 0$)]{ \includegraphics[width=0.45\columnwidth]{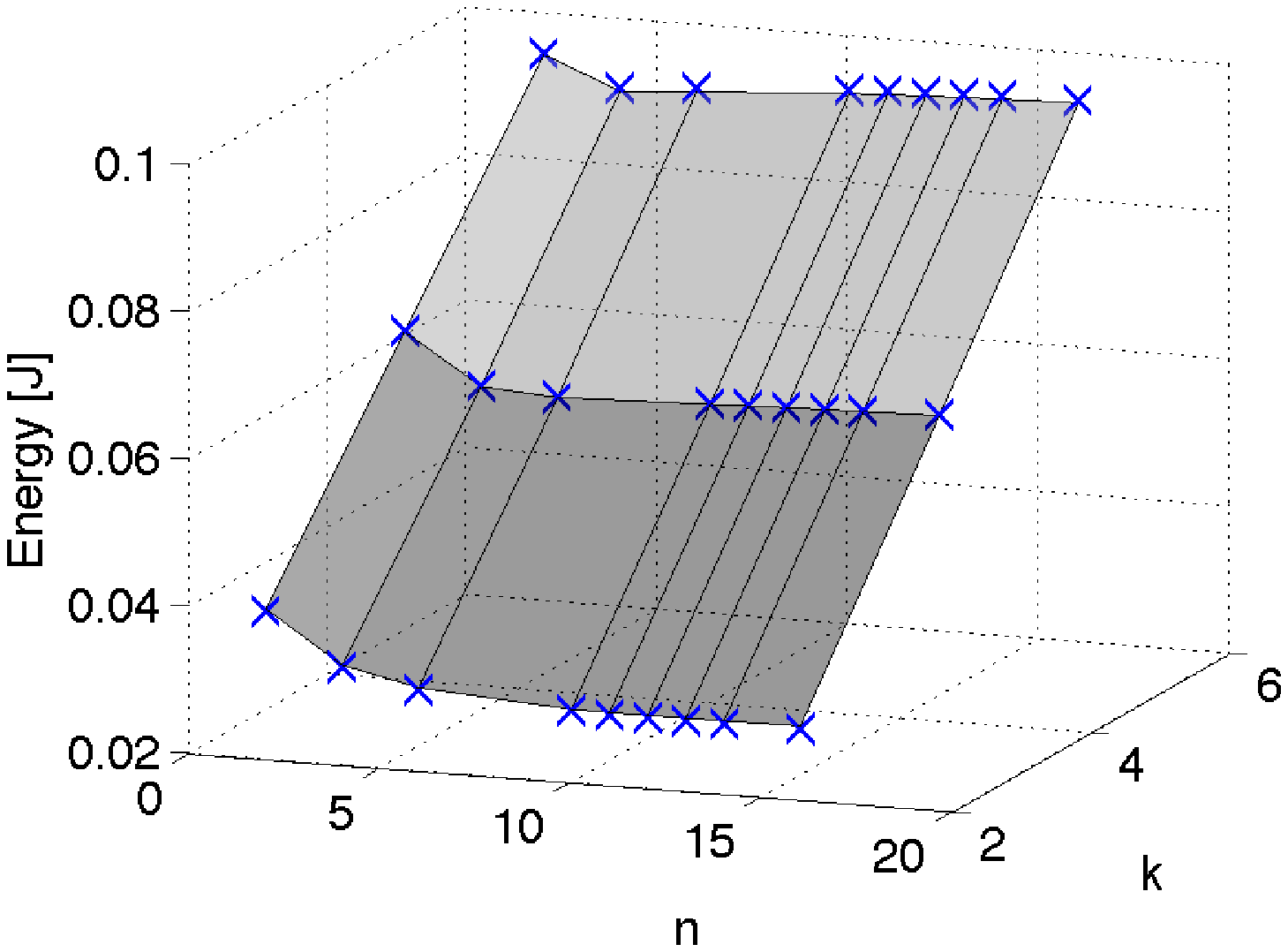}
\label{fig:uni-0} } \subfigure[Pareto PDF ($\alpha = 4$, $d = 0$)]{
\includegraphics[width=0.45\columnwidth]{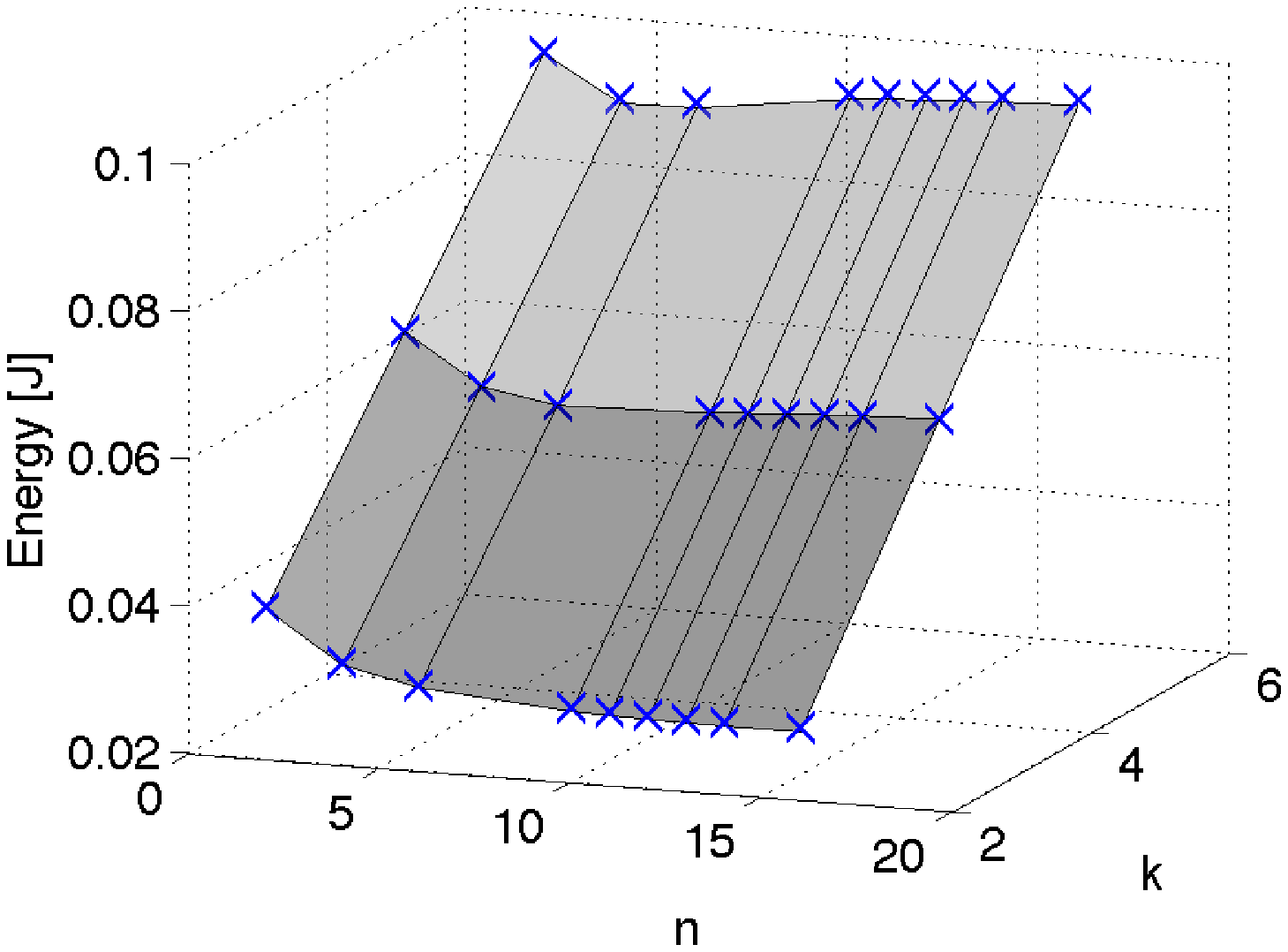}
\label{fig:par-0} } \subfigure[Exponential PDF ($d = 0$)]{ \includegraphics[width=0.45\columnwidth]{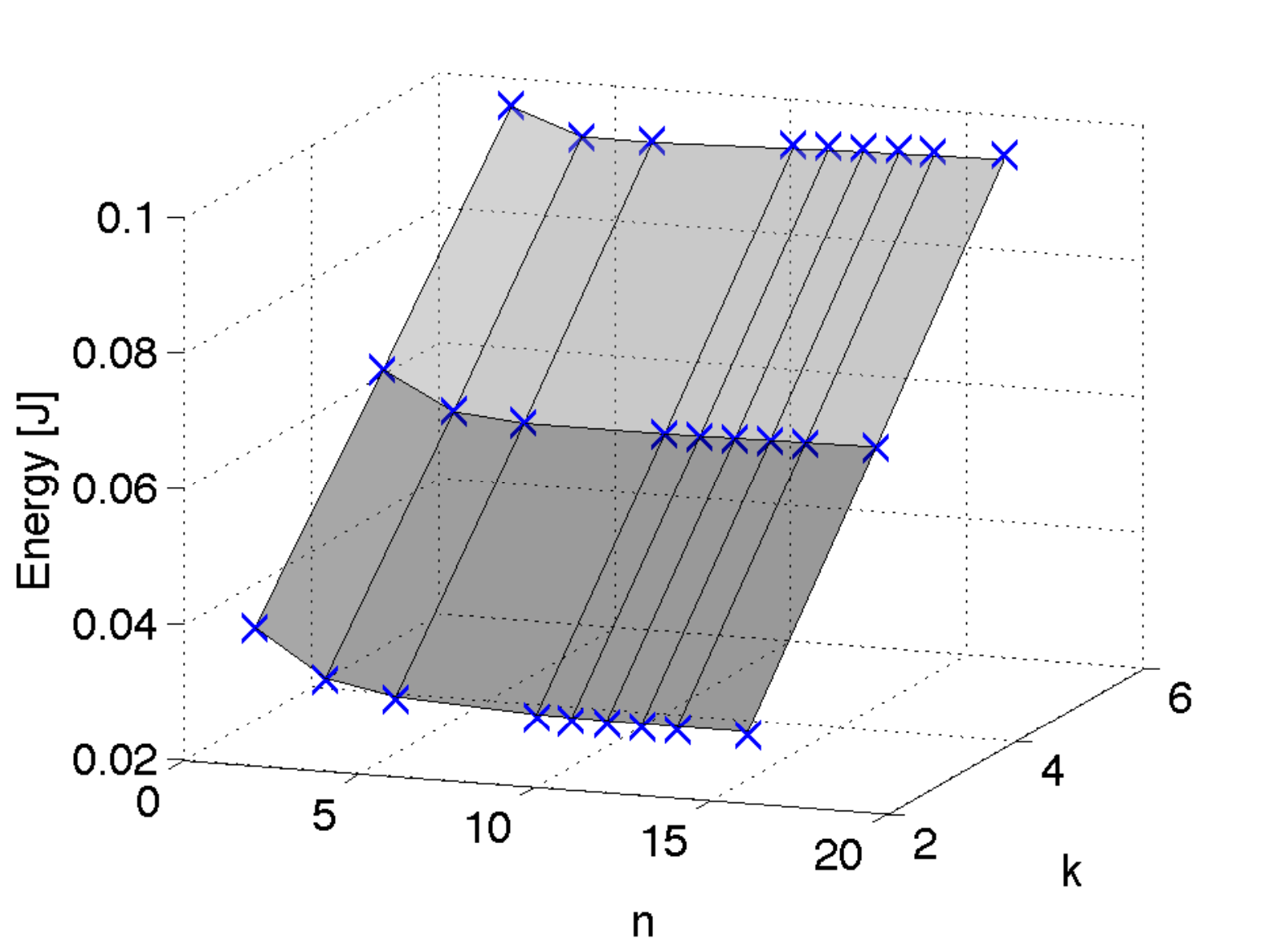}
\label{fig:exp-0} } \subfigure[Half-Gaussian PDF ($d = 0$)]{ \includegraphics[width=0.45\columnwidth]{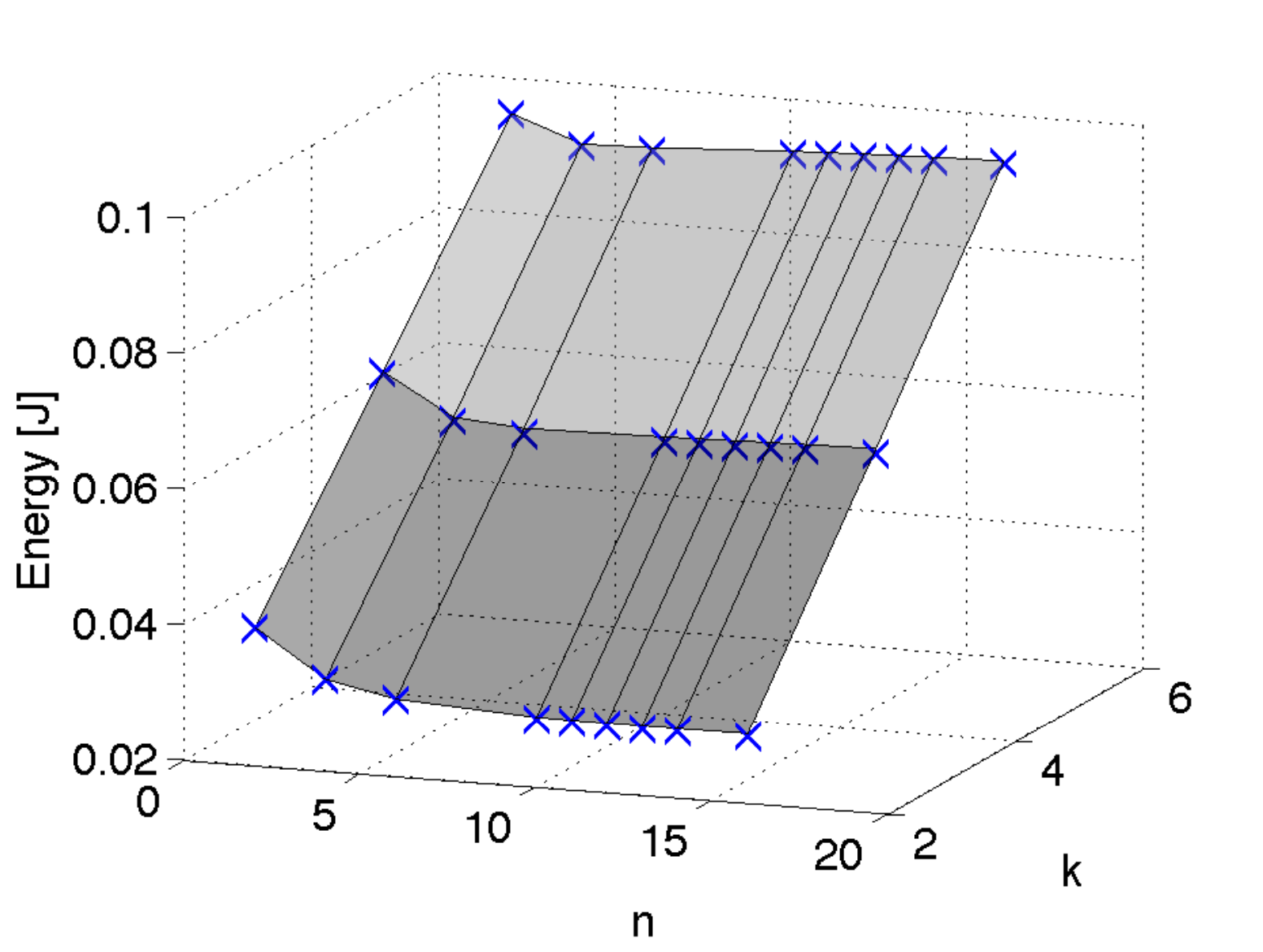}
\label{fig:hg-0} } \subfigure[Uniform PDF ($d = 2$)]{ \includegraphics[width=0.45\columnwidth]{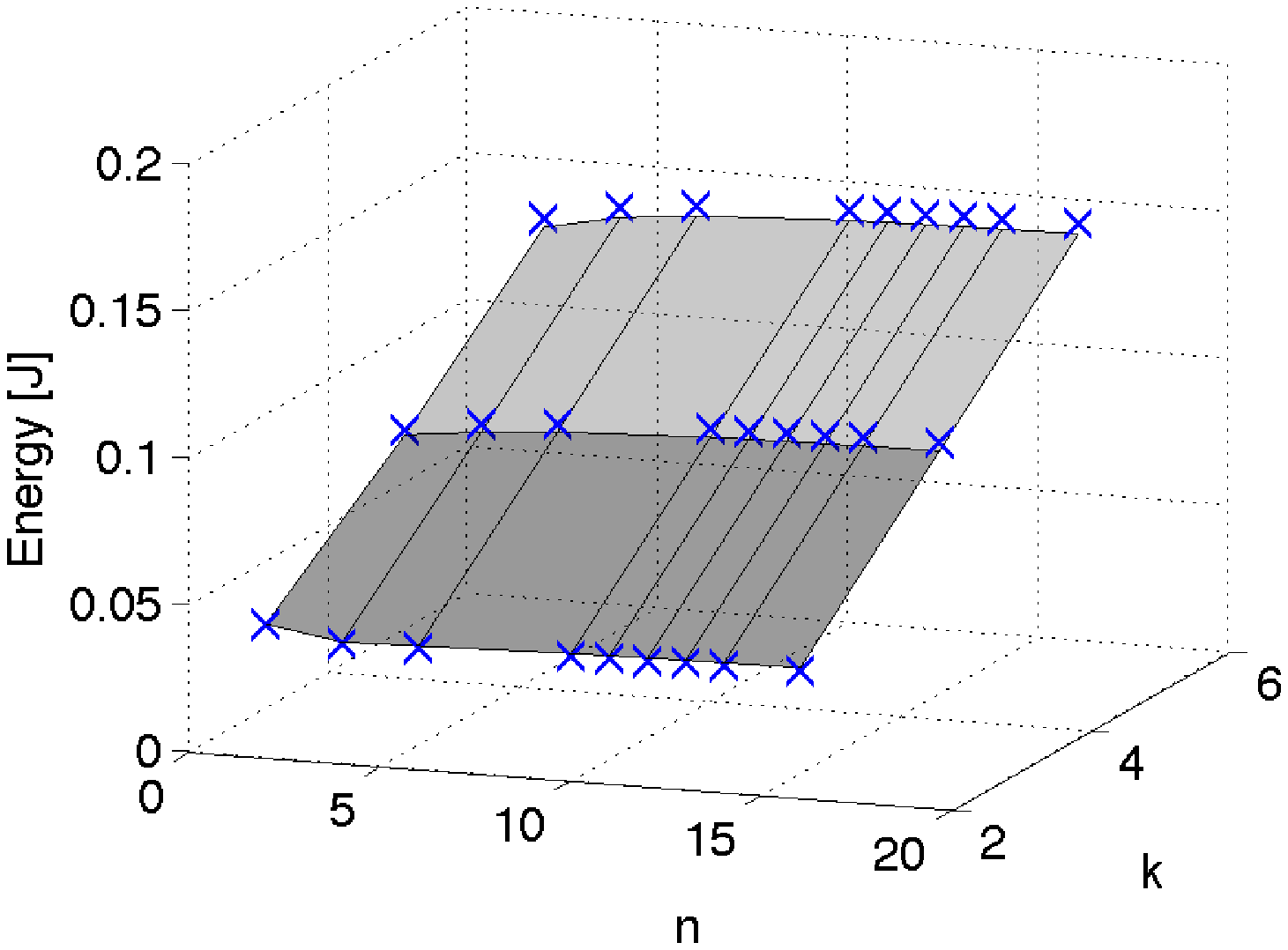}
\label{fig:uni-2} } \subfigure[Pareto PDF ($\alpha$ = 4, $d = 2$)]{
\includegraphics[width=0.45\columnwidth]{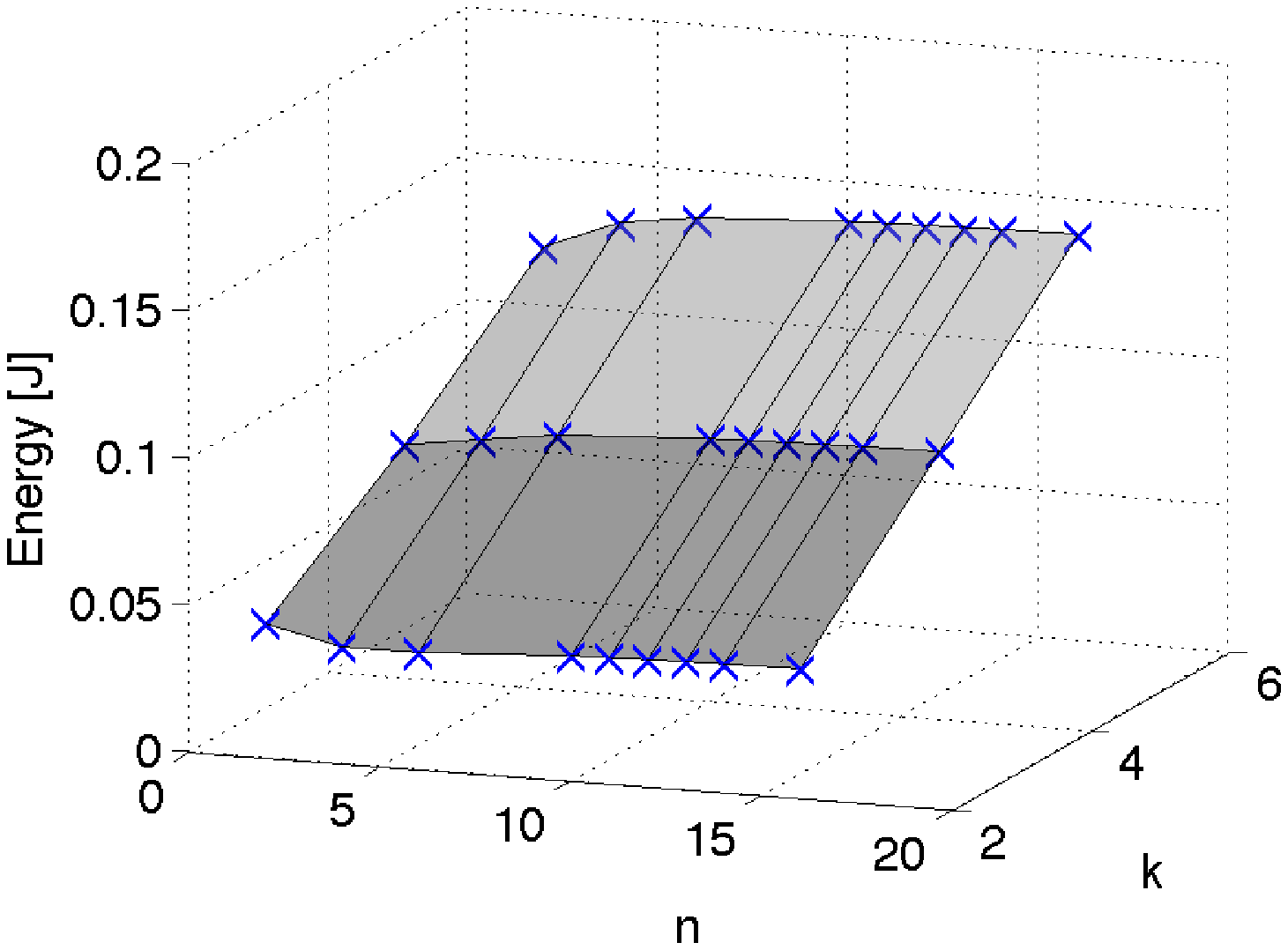}
\label{fig:par-2} } \subfigure[Exponential PDF ($d = 2$)]{ \includegraphics[width=0.45\columnwidth]{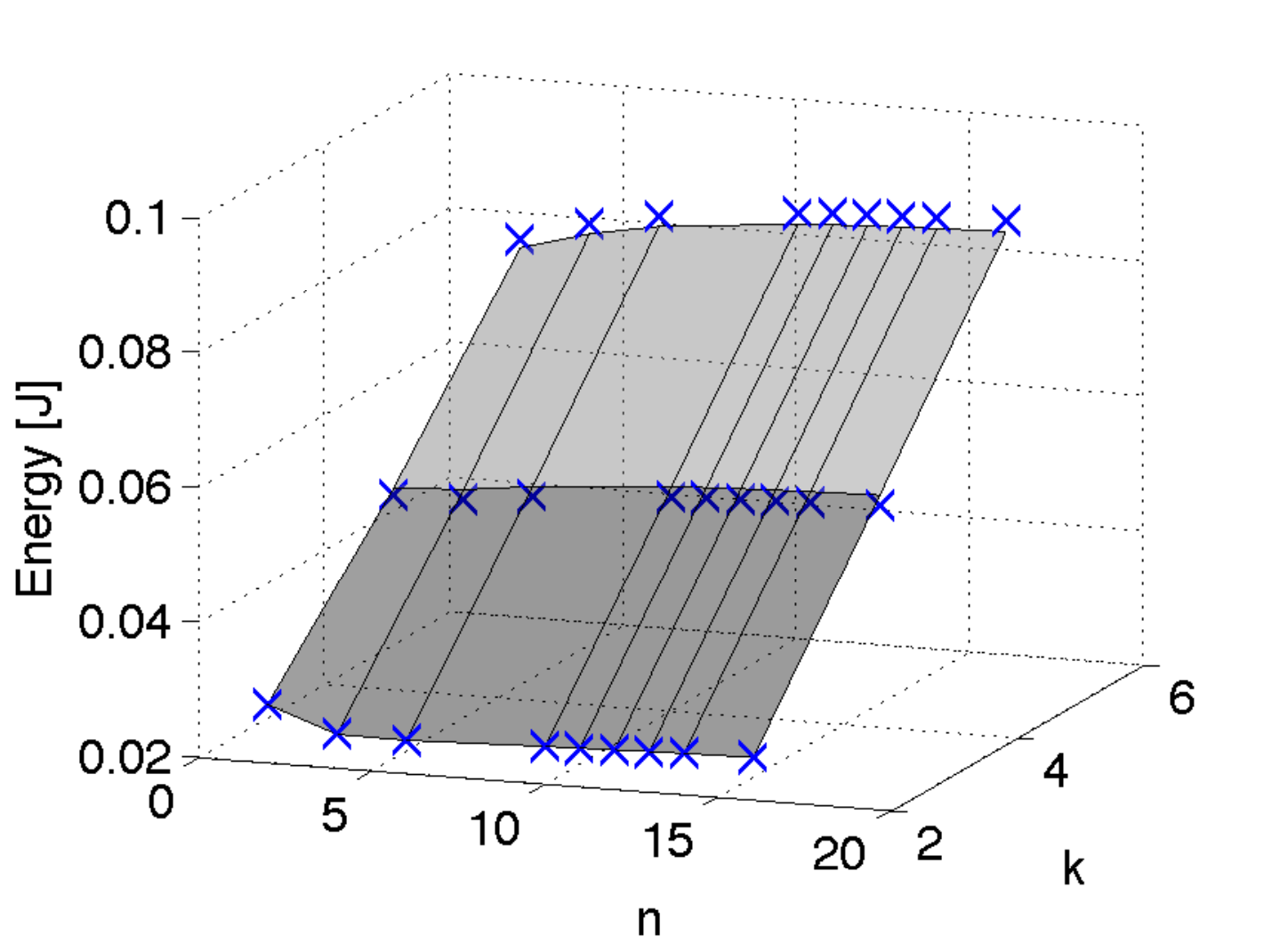}
\label{fig:exp-2} } \subfigure[Half-Gaussian PDF ($d = 2$)]{ \includegraphics[width=0.45\columnwidth]{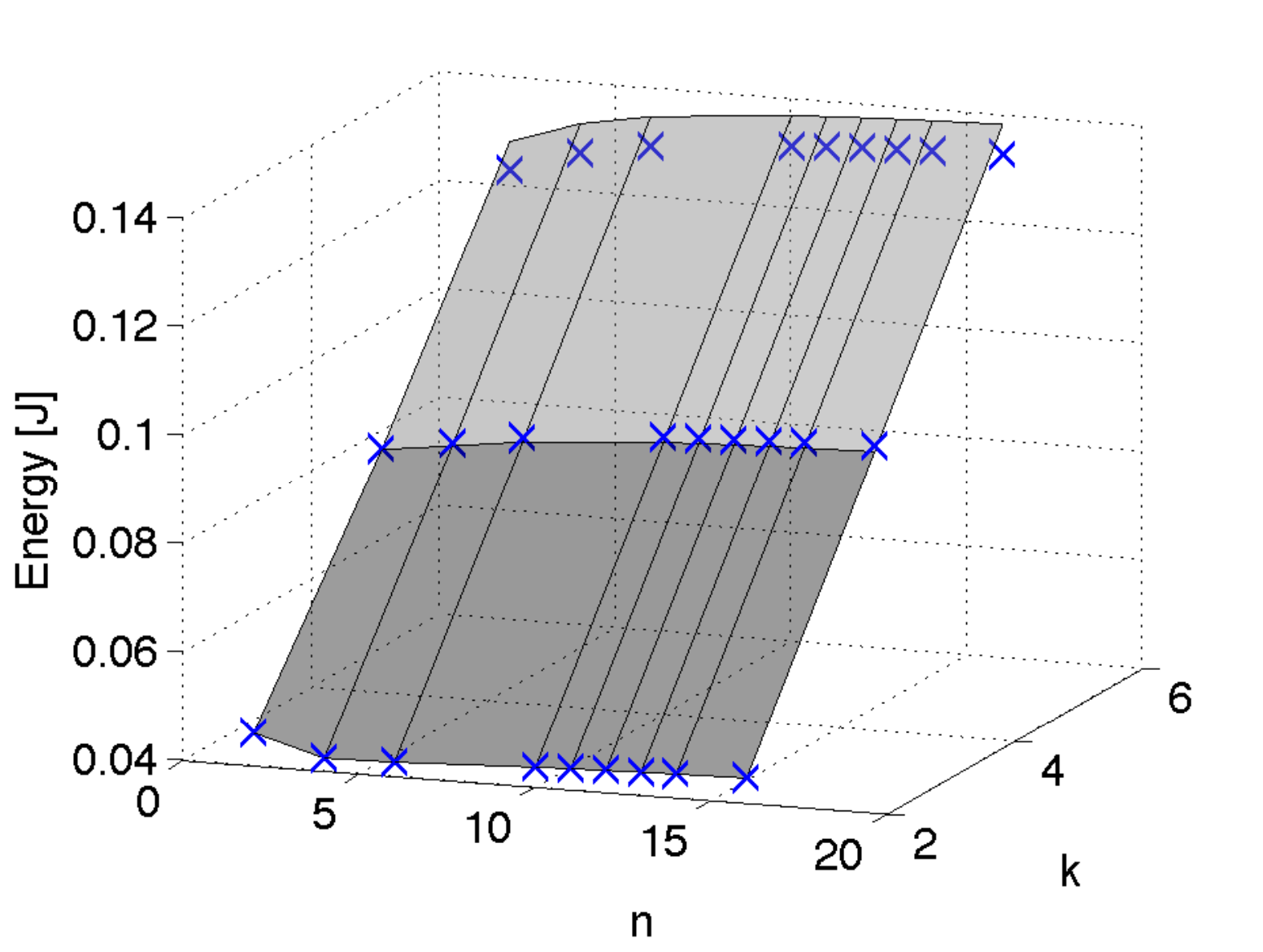}
\label{fig:hg-2} }

\caption{Each column shows the results corresponding to a marginal PDF characterizing
the data transmission process. The grayscale surfaces show the energy
consumption of a single camera sensor node in function of the number
of frames per second and the total number of nodes. The blue crosses
correspond to the value of the consumed energy as measured from the
sensor network testbed. All energy values and frames ($k$) are normalized
to an one-second interval. }

\label{fig:dist_all} 
\end{figure*}

\subsection{Model Validation via Monte-Carlo--generated Data }

\label{subsec:model_validation}

Under the settings described previously and shown in Table \ref{tab:TelosB Settings},
our first goal is to validate the analytic expressions of Section
\ref{sec:min_energy_analysis} that form the mathematical foundation
for Propositions 1 and 2, namely \eqref{eq:uniform_energy}, \eqref{eq:pareto_energy},
\eqref{eq:exponential_energy} and \eqref{eq:halfgaussian_energy}.
To this end, we create a controlled multimedia data production process
on each VSN node by: \emph{(i)} artificially creating several sets
of bitstream sizes according to the marginal PDFs of Section \ref{sec:min_energy_analysis}
via rejection sampling \cite{gilks1992adaptive}; \emph{(ii)} setting
the mean data size per video frame to $r=5.2$ kbit; \emph{(iii)}
setting $d=0$ (no relaying) and $d=2$ for each distribution. The
sets containing data sizes are copied onto the read-only memory of
each sensor node during deployment. At run time, each node fetches
a new frame size from the preloaded set, produces artificial data
according to it (akin to receiving the information from the multimedia
subsystem) and transmits the information to the LPBR following the
process described in the system model of Section \ref{sec:System Model}.
Depending on the frame size, the node can enter in idling/beaconing
state, or it can buffer the data exceeding the allocated TFDMA slots.
This controlled experiment with Monte-Carlo--generated datasets creates
the conditions that match our statistical characterization and can
therefore confirm the validity of our derivations. 

We report here energy measurements obtained under varying values of
$n$ and $k$. The chosen active time interval was set to be $T=154$
seconds and, beyond measuring the accuracy of the model versus experiments,
we also compared the theoretically-optimal values for $k$ and $n$
according to Section \ref{sec:min_energy_analysis} with the ones
producing the minimum energy consumption in the experiments. For the
reported experiments of Figures \ref{fig:dist_all}, and Table \ref{tab:Diff-theory&experiment-1-1},
the spatio--temporal constraints were: $N_{\min}=2$, $N_{\max}=16$
and $K_{\min}=2T$ frames, i.e. two frames per second. All our reported
measurements and the values for $k$ are normalized to a one-second
interval for easier interpretation of the results.

As one can see from Figures \ref{fig:dist_all}, and Table \ref{tab:Diff-theory&experiment-1-1},
the theoretical results match the experimental results for all the
tested distributions, with the maximum percentile error between them
limited to $6.34\%$ and all the coefficients of determination $R^{2}$
between the experimental and the model points being above $0.995$.
In addition, the theoretically-obtained optimal values for $\left\{ n^{\star},k^{\star}\right\} $
from \eqref{eq:uniform_optimal-1} and \eqref{eq:halfgaussian_optimal}
are always in agreement with the experimentally-derived values that
were found to offer the minimum energy consumption under the chosen
spatio--temporal constraints. We have observed the same level of accuracy
for the proposed model under a variety of data sizes ($r$), active
time interval durations ($T$), number of relay nodes ($d$) and spatio--temporal
constraints ($N_{\min}$, $N_{\max}$ and $K_{\text{min}}$), but
omit these repetitive experiments for brevity of exposition.

\begin{figure}
\begin{centering}
\includegraphics[width=0.7\columnwidth]{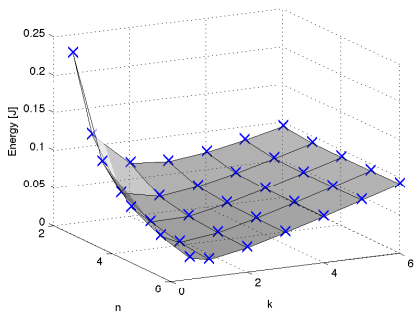}
\par\end{centering}

\caption{Predicted (gray surface) and measured (blue crosses) energy consumption
of a single camera node in function of the number of frames per second
and the total number of nodes, for the case of Uniform distribution
under $N_{\min}=2$, $N_{\max}=6$ and $K_{\min}=\frac{T}{2}$. All
energy values and frames ($k$) are normalized to an one-second interval.
\label{fig:dist_kmin}}
\end{figure}

As mentioned in Section \ref{sec:min_energy_analysis}-C, the optimal
solution does not always correspond to the minimum allowable number
of frames (i.e., $K_{\text{min}}$). For instance, Figure \ref{fig:dist_kmin}
shows the theoretical and experimental results obtained by setting
$N_{\min}=2$, $N_{\max}=6$ and $K_{\min}=\frac{T}{2}$ (i.e., one
frame every two seconds), and using the Uniform distribution. Under
these settings, the optimal solution was found to be $\left\langle n^{\star}=6,\, k^{\star}=T\right\rangle $,
thereby confirming the validity of the proposed model.

\begin{table*}
\begin{centering}
\caption{Differences between the theoretical and experimental results and the
optimal values, $\left\{ n^{\star},k^{\star}\right\} _{\text{D}}$,
for the number of nodes and the frames-per-second for the considered
data transmission (marginal) PDFs under the settings of Figure \ref{fig:dist_all}
and $d=2$ (each node relaying data from two other nodes).\label{tab:Diff-theory&experiment-1-1}}

\par\end{centering}

\centering{}%
\begin{tabular}{|c|c|c|c|c|c|c|c|c|}
\hline 
 & \multicolumn{4}{c|}{\textbf{$d=0$ (no relay transmission)}} & \multicolumn{4}{c|}{\textbf{$d=2$ (relaying from two other nodes)}}\tabularnewline
\hline 
\textbf{Transmission}  & \textbf{Mean}  & \textbf{Max.}  & \textbf{$R{}^{2}$}  & \textbf{Theoretical}  & \textbf{Mean}  & \textbf{Max.}  & \textbf{$R{}^{2}$}  & \textbf{Theoretical} \tabularnewline
\textbf{PDF}  & \textbf{error (\%)}  & \textbf{error (\%)}  & \textbf{coeff.}  & \textbf{optimum} & \textbf{error (\%)}  & \textbf{error (\%)}  & \textbf{coeff.}  & \textbf{optimum}\tabularnewline
\hline 
\hline 
Uniform  & 1.19  & 2.24  & 0.9982  & $\left\{ 12,2\right\} $  & 1.37  & 2.21  & 0.9921  & $\left\{ 4,2\right\} $ \tabularnewline
\hline 
Pareto $(\alpha=4)$  & 1.40  & 3.6  & 0.9980  & $\left\{ 16,2\right\} $  & 1.51  & 6.34  & 0.99983  & $\left\{ 6,2\right\} $ \tabularnewline
\hline 
Exponential  & 1.36  & 2.85  & 0.9984  & $\left\{ 15,2\right\} $  & 3.05  & 4.52  & 0.9895  & $\left\{ 5,2\right\} $ \tabularnewline
\hline 
Half-Gaussian  & 0.37  & 0.69  & 0.9991  & $\left\{ 13,2\right\} $  & 1.33  & 2.24  & 0.9977  & $\left\{ 4,2\right\} $ \tabularnewline
\hline 
\end{tabular}
\end{table*}

\section{Applications\label{sec:Applications}}

\begin{figure*}[t]
\centering\subfigure[DCT-DPCM coding, $R^2 = 0.9698$]{\includegraphics[width=0.7\columnwidth]{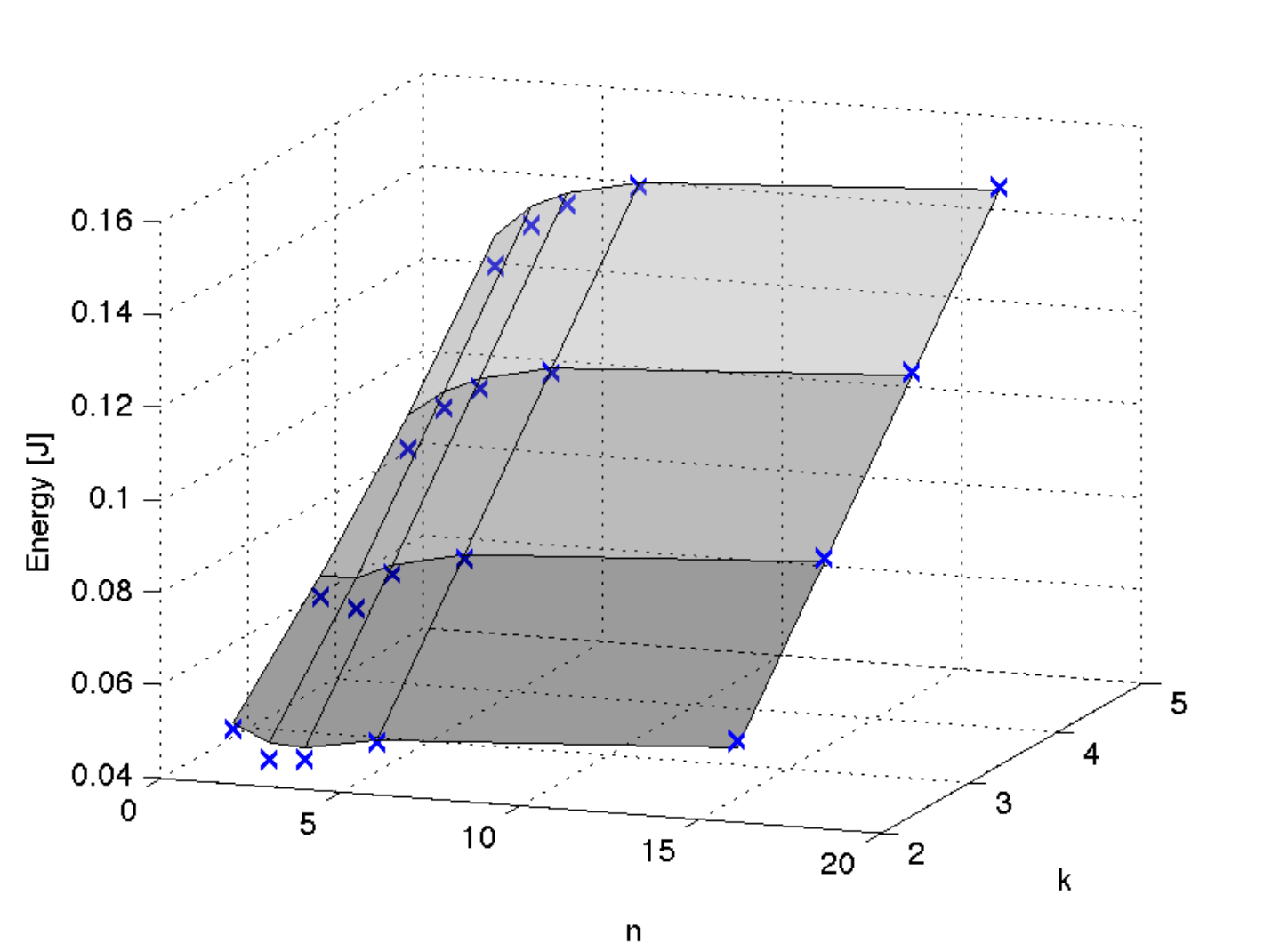}
\label{fig:Energy-Consumption-results-JPEG}} \subfigure[Visual features extraction, $R^2 = 0.9596$]{\includegraphics[width=0.7\columnwidth]{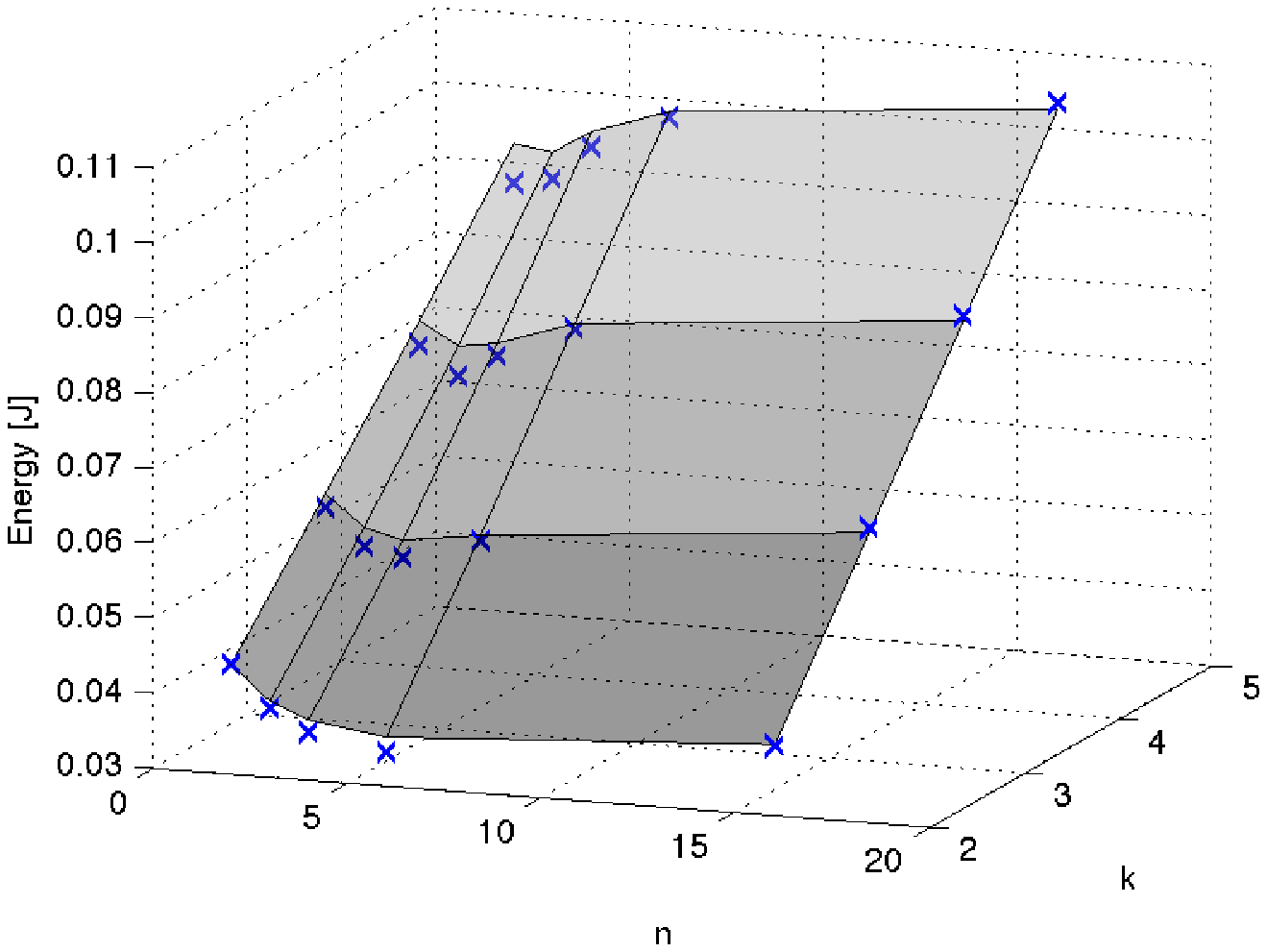}
\label{fig:Energy-Consumption-results-SP}} \caption{The energy function for the two considered application scenarios.
The grayscale surfaces represent the fitted energy function obtained
with the Pareto PDF, while the blue crosses represent the experimental
measurements. All energy values and frames ($k$) are normalized to
a one-second interval.}
\end{figure*}

In order to assess the proposed model against real application data,
we repeated the experimental measurements described in Section \ref{subsec:model_validation}
for both application scenarios and under the same spatio--temporal
constraints ($N_{\min}=2$, $N_{\max}=16$, $K_{\min}=2T$, i.e. two
frames per second), this time capturing and processing real data from
our deployment and utilizing the energy parameters of Table \ref{tab:TelosB Settings}
for the proposed analytic model. We then matched%
\footnote{Fitting is performed by matching the average data size $r$ of each
distribution to the average data size of the JPEG compressed frames
or the set of visual features.%
} the energy measurements with one of the energy functions derived
in Section \ref{sec:min_energy_analysis}. Specifically, we found
that the results matched best the Pareto distribution with parameters
$\alpha=4$, $v=kr$ and $r=20.6$ kbit for the JPEG case and $r=11.7$
kbit for the visual features case, as shown in Figures \ref{fig:Energy-Consumption-results-JPEG}
and \ref{fig:Energy-Consumption-results-SP}, with coefficient of
determination value $R^{2}\cong0.97$ for the JPEG case and $R^{2}\cong0.96$
for the visual features case. Similarly as before, all reported energy
values and number of frames are normalized to a one-second interval
for easier interpretation of the results.

Given the high accuracy of the Pareto-based energy model against the
application results, we utilized the settings for the minimum energy
consumption derived for the Pareto case {[}see \eqref{eq:uniform_optimal-1}{]}
to ascertain the energy saving that can be potentially achieved against
arbitrary (\emph{ad-hoc}) settings. As an example, in Tables \ref{tab:apps-jpeg}
and \ref{tab:apps-vf}, we consider two different cases for each application
scenario, characterized by different spatio--temporal constraints.
For each case, we compare the optimal solution given by \eqref{eq:uniform_optimal-1}
(for the Pareto case) with an\emph{ ad-hoc} ``least-cost'' solution
that assumes values equal to the minimum spatio--temporal constraints
(under the intuitive assumption that less nodes and less frames-per-second
lead to smaller energy consumption). Evidently, the proposed approach
allows for 8\% to 37\% energy savings in comparison to the \emph{ad-hoc}
settings in both applications under consideration. As such, its usage
can be envisaged for early-stage testing of plausible application
deployments with respect to their energy efficiency in order to determine
the impact of various options for the multimedia and radio subsystems,
as well as the best spatio--temporal parameters to consider, prior
to more detailed experimentation in the field.

\begin{table}
\caption{Motion JPEG application scenario \label{tab:apps-jpeg}}
\begin{tabular}{|c||c|c|c|}
\hline 
Constraints & \emph{Ad-hoc} deployment  & Proposed approach  & Gain \tabularnewline
\hline 
\hline 
$K_{\text{min}}=0.7$ & $k=0.7$  & $k=0.7$  & \tabularnewline
$N_{\text{min}}=2$  & $n=2$  & $n=10$  & 37.4\% \tabularnewline
$N_{\text{max}}=10$  & $E_{\text{c}}=0.027$ J  & $E_{\text{c}}=0.017$ J  & \tabularnewline
\hline 
\hline 
$K_{\text{min}}=2$ & $k=2$  & $k=2$  & \tabularnewline
$N_{\text{min}}=2$  & $n=2$  & $n=4$  & 7.9\%\tabularnewline
$N_{\text{max}}=10$  & $E_{\text{c}}=0.053$ J  & $E_{\text{c}}=0.049$ J  & \tabularnewline
\hline 
\end{tabular}
\end{table}

\begin{table}
\caption{Visual Features extraction application scenario \label{tab:apps-vf}}
\begin{tabular}{|c||c|c|c|}
\hline 
Constraints & \emph{Ad-hoc} deployment  & Proposed approach  & Gain \tabularnewline
\hline 
\hline 
$K_{\text{min}}=1.25$ & $k=1.25$  & $k=1.25$  & \tabularnewline
$N_{\text{min}}=2$  & $n=2$  & $n=10$  & 30.8\% \tabularnewline
$N_{\text{max}}=10$  & $E_{\text{c}}=0.033$ J  & $E_{\text{c}}=0.023$ J  & \tabularnewline
\hline 
\hline 
$K_{\text{min}}=2$ & $k=2$  & $k=2$  & \tabularnewline
$N_{\text{min}}=2$  & $n=2$  & $n=7$  & 18.1\%\tabularnewline
$N_{\text{max}}=10$  & $E_{\text{c}}=0.045$ J  & $E_{\text{c}}=0.037$ J  & \tabularnewline
\hline 
\end{tabular}
\end{table}

\section{Conclusions\label{sec:Conclusions}}

We proposed an analytic model for the energy consumption of a uniformly-formed
wireless visual sensor network (VSN) under varying spatio--temporal
constraints, defined in terms of number of nodes to be deployed per
network tier and video frames to be captured by each node. Analytic
conditions for the optimal spatio--temporal settings within the VSN
were derived for different probability density functions characterizing
the multimedia data volume to be transmitted by each node. Monte-Carlo
experiments performed via an energy-measurement testbed revealed that
the proposed model's accuracy is within 7\% of the obtained energy
consumption. Applying the model to two realistic scenarios for motion
JPEG compression and local visual features extraction within each
node in the VSN demonstrated that substantial energy savings can be
obtained via the proposed approach against \emph{ad-hoc} settings
for the spatio--temporal parameters of the VSN. As such, the proposed
model can be used for early-stage studies of VSNs to determine the
best operational parameters to be considered prior to cumbersome and
costly real-world deployment and testing.

\appendices{}

\section{\label{sec:Appendix-I}}

We first present the detailed proof of Proposition 1 under the Uniform
distribution ($\text{D}=\text{U}$). The proofs for the Pareto, Exponential
and Half-Gaussian distributions (i.e., Proposition 2) are summarized
afterward, since they follow the same steps as for the case of the
Uniform.

\subsection{Proof of Proposition 1 for the Uniform Distribution\label{sub:Appendix-Uniform}}

\subsubsection{Investigating the $n$-direction}

We examine the function $E_{\text{c,U}}$ along the plane $k=\bar{k}$,
$\bar{k}\ge K_{\text{min}}$, and analyze $E_{\text{c,U}}(n,\,\bar{k})$
which is now a function of $n$ only. It is straightforward to show
by first-derivative analysis that the only candidate extremum or inflection
point of $E_{\text{c,U}}(n,\,\bar{k})$ is $n_{\text{0,U}}=\frac{\beta_{\text{U}}}{\bar{k}}$,
with $\beta_{\text{U}}$ given by \ref{eq:beta_U}. This candidate
extremum holds under the assumption that: $N_{\textnormal{min}}\le n_{\text{0,U}}\le N_{\textnormal{max}}$,
i.e. that the candidate extremum or inflection point of $E_{\text{c,U}}(n,\,\bar{k})$
falls within the predefined spatial constraints of \eqref{eq:n_k_constraints}.
Furthermore, we find that $\left.\frac{d^{2}E_{\text{c,U}}(n,\,\bar{k})}{dn^{2}}\right|_{n=n_{\text{0,U}}}>0$,
which demonstrates that $n_{\text{0,U}}$ is a local minimum. Given
that local extrema must alternate within the region of support of
a continuous and differentiable function \cite{hardy2008course},
$n_{\text{0,U}}$ is also the global minimum of $E_{\text{c,U}}(n,\,\bar{k})$
within $N_{\textnormal{min}}\le n\le N_{\textnormal{max}}$.

Having derived the global minimum of $E_{\text{c,U}}(n,\,\bar{k})$
along an arbitrary plane $k=\bar{k}$, $\bar{k}\ge K_{\text{min}}$,
we can now attempt to find the value of $k$, $k\ge K_{\text{min}}$,
that minimizes the energy function. Evaluating $E_{\text{c,U}}(n,\, k)$
on $n=n_{\text{0,U}}$, we obtain: 
\begin{align}
E_{\text{c,U}}(n_{\text{0,U}},\, k) & =k\left[a+r\left[\left(p+j\right)\left(d+1\right)\right.\right.\nonumber \\
 & \left.\left.-\frac{p^{2}(d+1)}{b+p}+hd+g\right]\right].\label{eq:E_cU(n_0U,k)}
\end{align}
Evidently, the value of $k$ minimizing \eqref{eq:E_cU(n_0U,k)} is
the minimum allowable, i.e. $k=K_{\textrm{min}}$. Thus, the solution
minimizing \eqref{eq:arg_min_Ec} in the $n$-direction is $\text{\ensuremath{\mathcal{S}}}_{n_{0,\text{U}}}=\left(\frac{\beta_{\text{U}}}{K_{\text{min}}},\; K_{\text{min}}\right).$
This solution holds under the constraint:

\begin{equation}
N_{\textnormal{min}}\le\frac{\beta_{\text{U}}}{K_{\text{min}}}\le N_{\textnormal{max}}.\label{eq:constraint_S_n_oU}
\end{equation}

\subsubsection{Investigating the $k$-direction}

Similarly, we cut $E_{\text{c,U}}(n,\, k)$ along the plane $n=\bar{n}$,
$N_{\textnormal{min}}\le\bar{n}\le N_{\textnormal{max}}$, and minimize
$E_{\text{c,U}}(\bar{n},\, k)$ which is now a function of $k$ only.
Following the steps presented earlier, we can show by first and second
derivative analysis that the global minimum of $E_{\text{c,U}}(\bar{n},\, k)$
occurs at $k_{\text{0,U}}=\frac{\gamma_{\textrm{U}}}{\bar{n}}$, with
$\gamma_{\text{U}}$ given by \eqref{gamma_U}. This global minimum
holds under the assumption that $k_{\text{0,U}}\geq K_{\min}$, due
the predefined temporal constraint of \eqref{eq:n_k_constraints}.
Having derived the global minimum of $E_{\text{c,U}}(\bar{n},\, k)$
along an arbitrary plane $n=\bar{n}$, $N_{\textnormal{min}}\le\bar{n}\le N_{\textnormal{max}}$,
we can now attempt to find the value of $n$, $N_{\textnormal{min}}\le n\le N_{\textnormal{max}}$,
that minimizes the energy function. Evaluating $E_{\text{c,U}}(n,\, k)$
on $k=k_{\text{0,U}}$ we obtain:{} 
\begin{align}
E_{\text{c,U}}(n,\, k_{\text{0,U}}) & =\frac{1}{n}\left[\left[a+r\left[\left(p+j\right)\left(d+1\right)+hd+g\right]\right]\gamma_{\textrm{U}}\right.\label{eq:E_cU(n,k_0U)}\\
 & \left.-ps+\frac{s^{2}(b+p)}{4r(d+1)\gamma_{\textrm{U}}}\right].\nonumber 
\end{align}

\noindent Evidently, the value of $n$ minimizing \eqref{eq:E_cU(n,k_0U)}
is the maximum allowable, i.e. $n=N_{\textnormal{max}}$. Hence, the
solution when attempting to minimize \eqref{eq:E_cU(n,k_0U)} in the
$k$-direction under the constraints of \eqref{eq:n_k_constraints}
is $\text{\ensuremath{\mathcal{S}}}_{k_{0,\text{U}}}=\left(N_{\text{max}},\;\frac{\gamma_{\textrm{U}}}{N_{\text{max}}}\right)$
under the constraint: 
\begin{equation}
K_{\min}\leq\frac{\gamma_{\textrm{U}}}{N_{\text{max}}}.\label{eq:constraint_S_k_oU}
\end{equation}

\subsubsection{Uniqueness of solution and solution when \eqref{eq:constraint_S_n_oU}
and \eqref{eq:constraint_S_k_oU} do not hold}

So far, we have found two solutions minimizing the energy consumption
of each node: $\text{\ensuremath{\mathcal{S}}}_{n_{0,\text{U}}}$,
which minimizes the energy in the $n$-direction by appropriately
choosing the number of nodes to deploy (spatial resolution), and $\text{\ensuremath{\mathcal{S}}}_{k_{0,\text{U}}}$,
which minimizes the energy in the $k$-direction by appropriately
setting the optimal number of frames to capture (temporal resolution)
during the active time interval. However, the following issues arise: 

\begin{enumerate}[leftmargin=*] 
\item Both solutions are only applicable under constraints \eqref{eq:constraint_S_n_oU}
and \eqref{eq:constraint_S_k_oU}. Is it possible that \emph{both}
constraints are satisfied and, if so, then what is the best solution
for \eqref{eq:arg_min_Ec}? 

\item Conversely, if \emph{neither} of these two constraints is satisfied,
then what is the optimal solution for \eqref{eq:arg_min_Ec}? 

\end{enumerate}

It turns out that the answer to both questions can be derived based
on the value of the temporal constraint, $K_{\text{min}}$, as it
is clarified in the following analysis.

Starting from \eqref{eq:constraint_S_n_oU}, with a few straightforward
manipulations we reach $\frac{\beta_{\text{U}}}{N_{\max}}\le K_{\text{min}}\le\frac{\beta_{\text{U}}}{N_{\min}}$.
The second constraint for $K_{\text{min}}$ is provided by \eqref{eq:constraint_S_k_oU}.
It is now easy to prove that $\beta_{\text{U}}>\gamma_{\textrm{U}}$
(see Appendix \ref{subsec:ff}), which demonstrates that the constraints
of the two solutions are \emph{non-overlapping}, as the lower bound
of \eqref{eq:constraint_S_n_oU} is larger than the upper bound of
\eqref{eq:constraint_S_k_oU}. This answers the first question.

To address the second question, we have to analyze what happens when
$\frac{\gamma_{\textrm{U}}}{N_{\text{max}}}<K_{\text{min}}<\frac{\beta_{\text{U}}}{N_{\max}}$
or $K_{\text{min}}>\frac{\beta_{\text{U}}}{N_{\min}}$, as neither
of $\text{\ensuremath{\mathcal{S}}}_{n_{0,\text{U}}}$ and $\text{\ensuremath{\mathcal{S}}}_{k_{0,\text{U}}}$
are applicable in such cases. It is straightforward to show that $\frac{\partial E_{\text{c,U}}}{\partial n}$
and $\frac{\partial E_{\text{c,U}}}{\partial k}$ are never zero within
these intervals. Hence, the solution we are looking for must lie on
one of the two boundary points: $\left(N_{\text{min}},K_{\text{min}}\right)$
or $\left(N_{\text{max}},K_{\text{min}}\right)$.

Let us focus on the case of $\frac{\gamma_{\textrm{U}}}{N_{\text{max}}}<K_{\text{min}}<\frac{\beta_{\text{U}}}{N_{\max}}$
and evaluate $E_{\text{c,U}}(n,\, k)$ on the boundary plane $n=N_{\text{max}}$.
Since $E_{\text{c}}(N_{\text{max}},k)$ is monotonically increasing
for $k>\frac{\gamma_{\textrm{U}}}{N_{\text{max}}}$ the optimal point
is $k=K_{\text{min}}$, which leads to the solution $\text{\ensuremath{\mathcal{S}}}_{\max\min}=\left(N_{\text{max}},\; K_{\text{min}}\right)$.
Similarly, let us look at the $k$ direction by evaluating the energy
function on the $k=K_{\text{min}}$ plane. Now $n_{\text{0,U}}=\frac{\beta_{\text{U}}}{K_{\text{min}}}$
is larger than $N_{\text{max}}$ and is thus not admissible. Since
$E_{\text{c,U}}(n,\, K_{\text{min}})$ is decreasing for $n<n_{\text{0,U}}$,
the optimal point is $n=N_{\text{max}}$, which also leads to the
solution $\text{\ensuremath{\mathcal{S}}}_{\max\min}$. Finally, when
$K_{\text{min}}>\frac{\beta_{\text{U}}}{N_{\min}}$ , following a
similar analysis we reach that the optimal solution is $\text{\ensuremath{\mathcal{S}}}_{\min\min}=\left(N_{\text{min}},\; K_{\text{min}}\right)$.{} 

Summarizing, when the data transmitted by each VSN node follows the
Uniform distribution of \eqref{eq:uniform}, the set of solutions
giving the minimum energy consumption in \eqref{eq:arg_min_Ec} under
the spatio--temporal constraints of \eqref{eq:n_k_constraints} is
given by \eqref{eq:uniform_optimal-1}.

\subsection{Proof of Proposition 1 for the Pareto Distribution\label{sub:Appendix-Pareto}}

Considering the energy consumption for the Pareto distribution $E_{\text{c,P}}$
in \eqref{eq:pareto_energy}, we follow the derivative-based analysis
along each direction and join the obtained minima along with their
constraints.

\subsubsection{$n$-direction}

The partial derivative of $E_{\text{c,P}}$ with respect to $n$ (i.e.
under a plane $k=\bar{k}$ with $\bar{k}\geq K_{\min}$) is: 
\begin{equation}
\frac{\partial E_{\text{c,P}}}{\partial n}=-\frac{bs}{n^{2}}+\frac{s}{n^{2}}\left(b+p\right)\left(\frac{vn}{s}\right)^{\alpha}.
\end{equation}
The only solution for $\frac{\partial E_{\text{c,P}}}{\partial n}=0$
that can be admissible under the constraints of \eqref{eq:n_k_constraints}
is $n_{\text{0,P}}=\frac{\beta_{\text{P}}}{\bar{k}}$, with $\beta_{\text{P}}$
given by \eqref{eq:beta_P}. It is straightforward to show that $n_{\text{0,P}}$
corresponds to the global minimum of $E_{\text{c,P}}\left(n,\bar{k}\right)$.
Evaluating $E_{\text{c,P}}$ for $n_{\text{0,P}}$ leads to

\begin{align}
E_{\text{c,P}}(n_{\text{0,P}},\, k) & =\bar{k}\left[a+r\left[\left(j-b\right)\left(d+1\right)+hd+g\right.\right.\nonumber \\
 & \left.\left.+\left(d+1\right)\left(b\right)^{\frac{\alpha-1}{\mathbf{\alpha}}}\left(b+p\right)^{\frac{1}{\alpha}}\right]\right],
\end{align}

\noindent which attains its minimum value for the minimum allowable
$\bar{k}$, i.e. at point $\text{\ensuremath{\mathcal{S}}}_{n_{0,\text{P}}}=\left(\frac{\beta_{\text{P}}}{K_{\text{min}}},\; K_{\text{min}}\right)$.
Now we have to ensure that $N_{\text{min}}\le n_{\text{0,P}}\le N_{\text{max}}$,
which gives $\frac{\beta_{\text{P}}}{N_{\text{max}}}\le K_{\text{min}}\le\frac{\beta_{\text{P}}}{N_{\text{min}}}$.
As discussed for the Uniform case, for values of $K_{\text{min}}$
outside this range, the optimal solution comprises the border points
$\left(N_{\text{max}},\, K_{\text{min}}\right)$ or $\left(N_{\text{min}},\, K_{\text{min}}\right)$,
depending on temporal constraint.

\subsubsection{$k$-direction}

The partial derivative of $E_{\text{c,P}}$ with respect to $k$ (i.e.
under a plane $n=\bar{n}$ with $N_{\min}\leq\bar{n}\leq N_{\max}$)
is:

\begin{align}
\frac{\partial E_{\text{c,P}}}{\partial k} & =a+r\left[\left(j+p\right)\left(d+1\right)+hd+g\right]\nonumber \\
 & +r(b+p)(d+1)\\
 & \times\left[\left(\frac{\bar{n}}{s}\right)^{\alpha-1}\left(\frac{kr(\alpha-1)(d+1)}{\alpha}\right)^{\alpha-1}-1\right].\nonumber 
\end{align}

\noindent \begin{flushleft}
The only solution for $\frac{\partial E_{\text{c,P}}}{\partial k}=0$
that is admissible under the constraints of \eqref{eq:n_k_constraints}
is $k_{\text{0,P}}=\frac{\gamma_{\textrm{P}}}{\bar{n}}$, with $\gamma_{\textrm{P}}$
defined in \eqref{eq:gamma_P}. The first constraint imposed on $k_{0,\textrm{P}}$
is that it must be positive, which leads to 
\par\end{flushleft}

\noindent 
\begin{equation}
b>j+\frac{a}{r(d+1)}+\frac{hd+g}{d+1}.\label{eq:k_positive}
\end{equation}
The last equation indicates that the global minimum of $k_{\text{0,P}}$
holds only if the energy consumption during the idle state is greater
than the energy during transmission. While this is possible from a
mathematical point of view, the physical reality of wireless transceivers
does not allow for this case to manifest in a practical setting. We
also note that, beyond the constraint of \eqref{eq:k_positive}, the
global minimum of $k_{\text{0,P}}$ holds under the assumption that
$k_{\text{0,P}}\geq K_{\min}$ due the predefined temporal constraint
of \eqref{eq:n_k_constraints}.

\noindent Evaluating $E_{\text{c,P}}(n,\, k)$ on $k=k_{\text{0,P}}$,
we obtain

\begin{align}
E_{\text{c,P}}(n,\, k_{\text{0,P}}) & =\frac{b+p}{n}\left[s^{1-\alpha}\left(\frac{\beta_{\textrm{P}}r(\alpha-1)(d+1)}{\alpha}\right)^{\alpha}\right.\nonumber \\
 & \times\left.\left(\alpha-1\right)^{-1}-\beta_{\textrm{P}}r\left(d+1\right)\right]\label{eq:E_cP(n,k_0P)}\\
 & +\frac{bs}{n}+\frac{\gamma_{\textrm{P}}}{N_{\text{max}}}\frac{\beta_{\textrm{P}}\left[a+r\left[\left(j+p\right)\left(d+1\right)+hd+g\right]\right]}{n}\nonumber 
\end{align}

\noindent Evidently, for $\alpha>1$, the value of $n$ minimizing
\eqref{eq:E_cP(n,k_0P)} is the maximum allowable, i.e. $n=N_{\textnormal{max}}$.
Hence, the solution when attempting to minimize the energy consumption
function in the $k$-direction under the constraints of \eqref{eq:n_k_constraints}
is $\text{\ensuremath{\mathcal{S}}}_{k_{0,\text{P}}}=\left(N_{\text{max}},\;\frac{\gamma_{\textrm{P}}}{N_{\text{max}}}\right)$
under the constraint $K_{\min}\leq\frac{\gamma_{\textrm{P}}}{N_{\textrm{max}}}$.
It is now easy to prove that $\beta_{\text{P}}>\gamma_{\textrm{P}}$
(see Appendix \ref{subsec:ff2}), which demonstrates that the constraints
of the two solutions are \emph{non-overlapping}.

\subsection{Exponential Distribution\label{sub:Appendix-Exponential}}

The energy consumption in the case of Exponential distribution is
$E_{\text{c,E}}$, given by \eqref{eq:exponential_energy}. We follow
the derivative-based analysis along each direction and join together
the obtained minima along with their constraints.

\subsubsection{$n$-direction}

The partial derivative of $E_{\text{c,E}}$ with respect to $n$ (i.e.
under a plane $k=\bar{k}$ with $\bar{k}\geq K_{\min}$) is:

\begin{align}
\frac{\partial E_{\text{c},\text{E}}}{\partial n} & =-\frac{bs}{n^{2}}+\frac{s}{n^{2}}\left(b+p\right)\exp\left(-\frac{s}{nkr(d+1)}\right),
\end{align}
which, under the constraints of \ref{eq:n_k_constraints}, is equal
to zero for $n_{\text{0,E}}=\frac{\beta_{\text{E}}}{\bar{k}}$, with
$\beta_{\text{E}}$ given by \eqref{eq:beta_E}. It is straightforward
to show that $n_{\text{0,E}}$ corresponds to the global minimum of
$E_{\text{c,E}}\left(n,\bar{k}\right)$. Evaluating $E_{\text{c,E}}$
for $n_{\text{0,E}}$ leads to: 
\begin{align}
E_{\text{c,E}}(n_{\text{0,E}},\bar{k}) & =\bar{k}\left[a+r\left[j\left(d+1\right)+hd+g\right.\right.\nonumber \\
 & \left.\left.+b\left(d+1\right)\text{ln}\left(\frac{b+p}{b}\right)\right]\right],
\end{align}
which has its minimum value for the minimum allowable $\bar{k}$,
i.e. at point $\text{\ensuremath{\mathcal{S}}}_{n_{0,\text{E}}}=\left(\frac{\beta_{\text{E}}}{K_{\text{min}}},\; K_{\text{min}}\right)$.
Now we have to ensure that $N_{\text{min}}\le n_{\text{0,E}}\le N_{\text{max}}$,
which leads to $\frac{\beta_{\text{E}}}{N_{\text{max}}}\le K_{\text{min}}\le\frac{\beta_{\text{E}}}{N_{\text{min}}}$.
Again, for values of $K_{\text{min}}$ outside this range, the optimal
solution comprises the border points $\left(N_{\text{max}},\, K_{\text{min}}\right)$
or $\left(N_{\text{min}},\, K_{\text{min}}\right)$, depending on
temporal constraint.

\subsubsection{$k$-direction}

The partial derivative of $E_{\text{c,E}}$ with respect to $k$ (i.e.
under a plane $n=\bar{n}$ with $N_{\min}\leq\bar{n}\leq N_{\max}$)
is: 
\begin{equation}
\begin{array}{cc}
\frac{\partial E_{\text{c,E}}}{\partial k} & =\left[a+r\left[\left(p+j\right)\left(d+1\right)+hd+g\right]\right]\\
 & +\left(b+p\right)\left[r\left(d+1\right)\left(\exp\left(-\frac{s}{\bar{n}kr(d+1)}\right)-1\right)\right.\\
 & \left.+\frac{s}{\bar{n}k}\exp\left(-\frac{s}{\bar{n}kr(d+1)}\right)\right].
\end{array}
\end{equation}
The only solution for $\frac{\partial E_{\text{c,E}}}{\partial k}=0$
that may be admissible under the constraints of \eqref{eq:n_k_constraints}
is $k_{\text{0,E}}=\frac{\gamma_{\textrm{E}}}{\bar{n}}$, with $\gamma_{\textrm{E}}$
defined in \eqref{eq:gamma_E}.

\noindent The first constraint imposed on $k_{0,\textrm{E}}$ is that
it must be positive. That is, the product-log function should be smaller
than -1. This is true when the argument of the product-log function
is limited within $\left(-\frac{1}{\exp},0\right)$ \cite{corless1996lambertw}.
That is:

\begin{equation}
-\frac{1}{\exp}<\frac{a-r\left[\left(b-j\right)\left(d+1\right)-hd-g\right]}{\exp\times r(d+1)(b+p)}<0.\label{eq:constraint_k_0,E_positive}
\end{equation}

It is easy to verify that a necessary condition for \eqref{eq:constraint_k_0,E_positive}
to hold is \eqref{eq:k_positive}. Thus, similar to the Pareto case,
while the the global minimum of $k_{\text{0,E}}$ is in principle
possible, it is not expected to be encountered in a practical setup.
Beyond the constraint of \eqref{eq:constraint_k_0,E_positive}, the
global minimum of $k_{\text{0,E}}$ holds under the assumption that
$k_{\text{0,E}}\geq K_{\min}$ due the predefined temporal constraint
of \eqref{eq:n_k_constraints}.

Having derived the global minimum of $E_{\text{c,E}}(\bar{n},\, k)$
along an arbitrary plane $n=\bar{n}$, $N_{\textnormal{min}}\le\bar{n}\le N_{\textnormal{max}}$,
we can now attempt to find the value of $n$, $N_{\textnormal{min}}\le n\le N_{\textnormal{max}}$,
that minimizes the energy function. Evaluating $E_{\text{c,E}}(n,\, k)$
on $k=k_{\text{0,P}}$ we obtain

\begin{align}
E_{\text{c,E}}(n,\, k_{\text{0,P}}) & =\frac{r\gamma_{\textrm{E}}(b+p)(d+1)}{n}\left(\exp\left(-\frac{1}{r\gamma_{\textrm{E}}(d+1)}\right)-1\right)\nonumber \\
 & +\frac{bs}{n}+\frac{\gamma_{\textrm{E}}(a+r\left[\left(p+j\right)\left(d+1\right)+hd+g\right]}{n}.\label{eq:Ec_E(n,k0_E)}
\end{align}

\noindent Evidently, for $\alpha>1$, the value of $n$ minimizing
\eqref{eq:Ec_E(n,k0_E)} is the maximum allowable, i.e. $n=N_{\textnormal{max}}$.
Hence, the solution when attempting to minimize the energy consumption
function in the $k$-direction under the constraints of \eqref{eq:n_k_constraints}
is $\text{\ensuremath{\mathcal{S}}}_{k_{0,\text{P}}}=\left(N_{\text{max}},\;\frac{\gamma_{\textrm{E}}}{N_{\text{max}}}\right)$
under the constraint $K_{\min}\leq\frac{\gamma_{\textrm{E}}}{N_{\textrm{max}}}$.
It is now easy to prove that $\beta_{\text{E}}>\gamma_{\textrm{E}}$
(see Appendix \ref{subsec:ff2}), which demonstrates that the constraints
of the two solutions are \emph{non-overlapping}.

\subsection{Half-Gaussian Distribution\label{sub:Appendix-Half-Gaussian}}

The energy consumption for half-Gaussian distribution is $E_{\text{c,H}}$
given by \eqref{eq:halfgaussian_energy}.

\subsubsection{$n$-direction}

The partial derivative of $E_{\text{c,H}}$ with respect to $n$ (i.e.
under a plane $k=\bar{k}$ with $\bar{k}\geq K_{\min}$) is: 
\begin{equation}
\frac{\partial E_{\text{c,H}}}{\partial n}=\frac{ps}{n^{2}}-\frac{s\left(b+p\right)}{n^{2}}\text{erf}\left(\frac{s}{\sqrt{\pi}\bar{k}rn(d+1)}\right),
\end{equation}
which, under the constraints of \ref{eq:n_k_constraints}, is equal
to zero for $n_{\text{0,H}}=\frac{\beta_{\text{H}}}{\bar{k}}$, with
$\beta_{\text{H}}$ given by \eqref{eq:beta_H}. It is easy to show
that $n_{\text{0,H}}$ corresponds to the global minimum of $E_{\text{c,H}}\left(n,\bar{k}\right)$.
Evaluating $E_{\text{c,H}}$ for $n_{\text{0,H}}$ leads to:

\begin{align}
E_{\text{c,H}}(n_{\text{0,H}},\, k) & =k\left[a+r\left[\left(p+j\right)\left(d+1\right)+hd+g\right.\right.\nonumber \\
 & +\left.\left.\left(b+p\right)\left(d+1\right)\exp\left(\left[-\text{ erf}^{-1}\left(\frac{p}{b+p}\right)\right]^{2}\right)\right]\right].
\end{align}
which has its minimum value for the minimum allowable $\bar{k}$,
i.e. at $\text{\ensuremath{\mathcal{S}}}_{n_{0,\text{H}}}=\left(\frac{\beta_{\text{H}}}{K_{\text{min}}},\; K_{\text{min}}\right)$.
Now, we have to ensure that $N_{\text{min}}\le n_{\text{0,H}}\le N_{\text{max}}$,
which leads to $\frac{\beta_{\text{H}}}{N_{\text{max}}}\le K_{\text{min}}\le\frac{\beta_{\text{H}}}{N_{\text{min}}}$.
Similarly as for the previous distributions, for values of $K_{\text{min}}$
outside this range, the optimal solution comprises the border points
$\left(N_{\text{max}},\, K_{\text{min}}\right)$ or $\left(N_{\text{min}},\, K_{\text{min}}\right)$.

\subsubsection{$k$-direction}

The partial derivative of $E_{\text{c,H}}$ with respect to $k$ (i.e.
under a plane $n=\bar{n}$ with $N_{\min}\leq\bar{n}\leq N_{\max}$)
is: 
\begin{align}
\frac{\partial E_{\text{c,H}}}{\partial k} & =\left[a+r\left[\left(p+j\right)\left(d+1\right)+hd+g\right]\right.\nonumber \\
 & +r\left(b+p\right)(d+1)\\
 & \times\left(\exp\left(-\frac{s^{2}}{\pi k^{2}r^{2}\bar{n}^{2}(d+1)^{2}}\right)-1\right),\nonumber 
\end{align}
which can be shown to be positive. Hence, the energy function is increasing
with respect to $k$ and the optimal value is the minimum allowable
$k$. Thus, the solution is equal to $\text{\ensuremath{\mathcal{S}}}_{n_{0,\text{H}}}$.

\section{\label{sec:Appendix-II}}

\subsection{Proof that $\beta_{\text{U}}>\gamma_{\textrm{U}}$}

\label{subsec:ff} Replacing $\beta_{\textrm{U}}$ and $\gamma_{\text{U}}$
from \eqref{eq:beta_U} and \eqref{gamma_U} in the inequality we
desire to prove, squaring both sides (since all terms are positive)
and rearranging terms, leads to 
\begin{align}
r\left[b\left(g+j+p\right)+p\left(g+j+pd\right)+bd\left(h+j+p\right)\right.\nonumber \\
\left.+pd\left(h+j\right)\right]+a\left(b+p\right) & >0,
\end{align}
which is indeed positive because all constants are positive quantities.

\subsection{Proof that $\beta_{\textrm{P}}>\gamma_{\textrm{P}}$ }

\label{subsec:ff2}Replacing the terms $\beta_{\textrm{P}}$ and $\gamma_{\textrm{P}}$
from \eqref{eq:gamma_P} and \eqref{eq:beta_P} in the inequality
we desire to prove, we reach:

\begin{equation}
\left(\frac{b}{b+p}\right)^{\frac{1}{\alpha}}>\left(\frac{-a+r\left[\left(b-j\right)\left(d+1\right)-hd-g\right]}{r\left(d+1\right)\left(b+p\right)}\right)^{\frac{1}{\alpha-1}}.\label{eq:bp_greater_gammap}
\end{equation}

\noindent Now, recalling the constraint of \eqref{eq:k_positive},
let us assume the minimum possible value for $b$, i.e.,

\begin{equation}
b=j+\frac{a}{r(d+1)}+\frac{hd+g}{d+1}+\delta,\label{eq:b_delta}
\end{equation}

\noindent with $\delta>0$. Evidently, $b>\delta$ since all constants
are positive. Substituting $b$ in the numerator of the right hand
side of \eqref{eq:bp_greater_gammap} via \eqref{eq:b_delta}, we
obtain $\left(\frac{b}{b+p}\right)^{\frac{1}{\alpha}}>\left(\frac{\delta}{b+p}\right)^{\frac{1}{\alpha-1}}$.
Since $b>\delta$, in order to prove the last expression it suffices
to prove that $\left(\frac{b}{b+p}\right)^{\frac{1}{\alpha}}>\left(\frac{b}{b+p}\right)^{\frac{1}{\alpha-1}}$
holds. The last expression is indeed true because $\frac{b}{b+p}\leq1$.

\subsection{Proof that $\beta_{\text{E}}>\gamma_{\textrm{E}}$ }

\label{subsec:ff3}Replacing $\beta_{\textrm{E}}$ and $\gamma_{\text{E}}$
from \eqref{eq:gamma_E} and \eqref{eq:beta_E} in the inequality
we desire to prove, we reach

\noindent 
\begin{align}
\frac{1}{\ln\left(\frac{b+p}{p}\right)} & >-\frac{1}{W\left(\frac{a-r\left[\left(b-j\right)\left(d+1\right)-hd-g\right]}{\exp\times r\left(d+1\right)\left(b+p\right)}\right)+1}.
\end{align}
Recalling that, under the constraint \eqref{eq:k_positive}, the Lambert
$W$ function is upper-bounded by -1 we obtain

\noindent 
\begin{equation}
-\ln\left(\frac{b+p}{p}\right)-1>W\left(\frac{a-r\left[\left(b-j\right)\left(d+1\right)-hd-g\right]}{\exp\times r\left(d+1\right)\left(b+p\right)}\right).\label{eq:be_greater_gammae}
\end{equation}

\noindent Substituting $b$ in the numerator of the right side of
\eqref{eq:be_greater_gammae} with the expression of \eqref{eq:b_delta}
and using the definition of the product-log function, $z=W\left(z\right)\exp\left(W\left(z\right)\right)$,
the last inequality leads to $p>\frac{-\delta}{W\left(\frac{-\delta}{\exp\times(b+p)}\right)}$.
The right-hand side is upper bounded by $\delta$, since the Lambert
function is upper bounded by -1. Thus, to complete the proof, it suffices
to prove that $p>\delta$. For derivating the solutions in the Exponential
case, we have assumed that $p>b$ and \eqref{eq:b_delta} shows that
$b>\delta$. Therefore, $p>\delta$.{} 

\balance

        \newpage
        \nobalance
         \begin{biography}[{\includegraphics[width=1in,height=1.25in,clip,keepaspectratio]{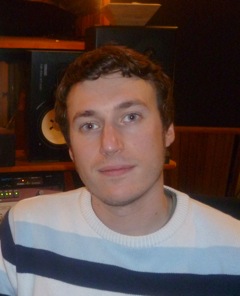}}]{Alessandro Redondi}
                received his Master degree in Computer Engineering in July 2009 and the Ph.D. in Information Engineering in 2014, both from Politecnico di Milano. Currently he is a post-doctoral researcher at the Department of Electronics and Information of Politecnico di Milano. His research activities are focused on algorithms and protocols for Visual Sensor Networks and Real Time Localization Systems.
                \end{biography}
                
\vfill
\vfill
\vfill
\vfill
\vfill
\vfill  
                                
                                \begin{biography}[{\includegraphics[width=1in,height=1.25in,clip,keepaspectratio]{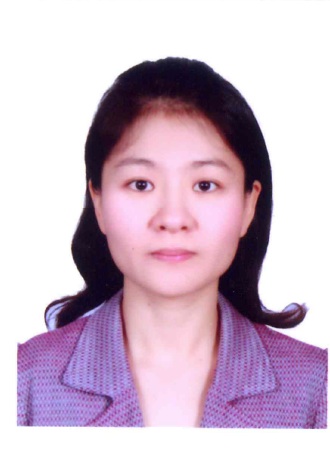}}]{Dujdow Buranapanichkit}
                                is Lecturer in the Department of Electrical Engineering, Faculty of Engineering, Prince of Songkla University (Thailand). She received her PhD from the Department of Electronic and Electrical Engineering of University College London (UK). Her research interests are in wireless sensor networks and distributed synchronization mechanisms and protocol design.
                                \end{biography}
\vfill
\vfill
\vfill
\vfill
\vfill
\vfill          

                \begin{biography}[{\includegraphics[width=1in,height=1.25in,clip,keepaspectratio]{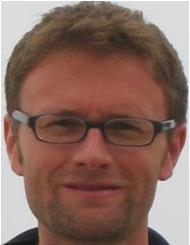}}]{Matteo Cesana}
                is currently an Assistant Professor with the Dipartimento di Elettronica, Informazione e Bioingegneria of the Politecnico di Milano, Italy. He received his MS degree in Telecommunications Engineering and his Ph.D. degree in Information Engineering from Politecnico di Milano in July 2000 and in September 2004, respectively. From September 2002 to March 2003 he was a visiting researcher at the Computer Science Department of the University of California in Los Angeles (UCLA). His research activities are in the field of design, optimization and performance evaluation of wireless networks with a specific focus on wireless sensor networks and cognitive radio networks. Dr. Cesana is an Associate Editor of the Ad Hoc Networks Journal (Elsevier).
                \end{biography}

                
\vfill
\vfill
\vfill
\vfill
\vfill
\vfill  

                \begin{biography}[{\includegraphics[width=1in,height=1.25in,clip,keepaspectratio]{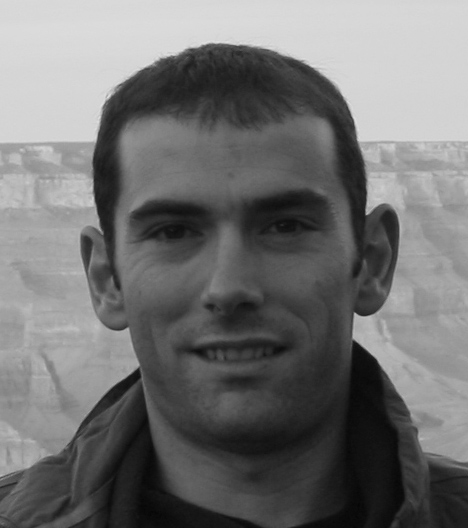}}]{Marco Tagliasacchi}
                        is currently Assistant Professor at the ``Dipartimento di Elettronica e Informazione - Politecnico di Milano'', Italy. He received the ``Laurea'' degree (2002, cum Laude) in Computer Engineering and the Ph.D. in Electrical Engineering and Computer Science (2006), both from Politecnico di Milano. He was visiting academic at the Imperial College London (2012) and visiting scholar at the University of California, Berkeley (2004).
His research interests include multimedia forensics, multimedia communications (visual sensor networks, coding, quality assessment) and information retrieval. Dr. Tagliasacchi co-authored more than 120 papers in international journals and conferences, including award winning papers at MMSP 2013, MMSP2012, ICIP 2011, MMSP 2009 and QoMex 2009. He has been actively involved in several EU-funded research projects. He is currently co-coordinating two ICT-FP7 FET-Open projects (GreenEyes - www.greeneyesproject.eu, REWIND - www.rewindproject.eu).
Dr. Tagliasacchi is an elected member of the IEEE Information Forensics and Security Technical committee for the term 2014-2016, and served as member of the IEEE MMSP Technical Committee for the term 2009-2012. He is currently Associate Editor for the IEEE Transactions on Circuits and Systems for Video Technologies (2011 best AE award) and APSIPA Transactions on Signal and Information Processing. Dr. Tagliasacchi was General co-Chair of IEEE Workshop on Multimedia Signal Processing (MMSP 2013, Pula, Italy) and he will be Technical Program Coordinator of IEEE International Conference on Multimedia\&Expo (ICME 2015, Turin, Italy).
                \end{biography}
                
                                                                \begin{biography}[{\includegraphics[width=1in,height=1.25in,clip,keepaspectratio]{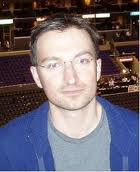}}]{Yiannis Andreopoulos} (M00) is Senior Lecturer in the Department of Electronic and Electrical Engineering of University College London (UK). His research interests are in wireless sensor networks, error-tolerant computing and multimedia systems.
                \end{biography}

\end{document}